\documentclass[ALICE,manyauthors]{cernphprep}

\RequirePackage{lineno}
\newcommand{\jpsi}{\rm J/$\psi$}
\newcommand{\pt}{\rm $p_{\rm t}$}

\newcommand{\dd}{\mathrm{d}} 
\newcommand{\chic}{$\chi_{\rm c}$}
\newcommand{\hb}{${\rm h_B}$}

\usepackage{epsfig}
\usepackage{epstopdf}
\usepackage{hyperref}
\usepackage{rotating,color,cite}

\begin{document}%
%
%
\begin{titlepage}
\PHnumber{2012-132}      
\PHdate{24 October 2012}             
%
%
\title{Measurement of prompt J/\boldmath${\psi}$ and beauty hadron production cross sections
at mid-rapidity in pp collisions at \boldmath$\sqrt{s}$~=~7~TeV}
\ShortTitle{Prompt J/$\mathbf{\psi}$ and beauty hadron production
at mid-rapidity  
in pp collisions at $\mathbf{\sqrt{s}}$=7 TeV}   
%
\Collaboration{The ALICE Collaboration%
         \thanks{See Appendix~\ref{app:collab} for the list of collaboration 
                      members}}
\ShortAuthor{The ALICE Collaboration}      
\begin{abstract}
 The ALICE experiment at the LHC
  has studied  \jpsi\ production at 
  mid-rapidity in pp collisions at $\sqrt{s}$~=~7~TeV 
  through its electron pair decay
  on a data sample corresponding to an integrated luminosity $L_{\rm int} = 5.6\, {\rm nb^{-1}}$.
  The fraction of \jpsi\ from the decay of long-lived beauty hadrons was determined
  for \jpsi\ candidates with transverse momentum $p_{\rm t}>1.3$~GeV/$c$ and rapidity $|y|<0.9$.
  The cross section for prompt \jpsi\ mesons, i.e. directly produced \jpsi\ and prompt decays
  of heavier charmonium states such as
  the ${\rm \psi(2S)}$ and ${\rm \chi_c}$ resonances,
  is $\sigma_{\rm prompt \, J/\psi} \left( p_{\rm t}> 1.3\;{\rm GeV}/c,\; |y|<0.9 \right)$ =
     $8.3\pm 0.8 \, {\rm (stat.)}\, \pm 1.1 \, {\rm (syst.)} \, ^{+ 1.5}_{-1.4} \, {\rm (syst. \, pol.) \; \mu b}$. 
  The 
  cross section for the production of b-hadrons 
  decaying to \jpsi\
  with $p_{\rm t}>1.3$~GeV/$c$ and $|y|<0.9$
  is
  $\sigma_{J/\psi \leftarrow {\rm h_B}} \left( p_{\rm t}> 1.3\;{\rm GeV}/c,\; |y|<0.9 \right) =
    1.46$ $\pm$~0.38~(stat.)~$^{+ 0.26}_{-0.32}$~(syst.)~$\mu$b.  
  The results are compared to QCD model predictions.
  The shape of the \pt\ and $y$ distributions of b-quarks
  predicted by
  perturbative  QCD  model calculations are used
  to extrapolate
  the measured cross section to derive
  the  ${\rm b \overline{b}}$ pair total cross section and
  $\dd \sigma / \dd y$ at mid-rapidity.  
\end{abstract}
\end{titlepage}
\section{Introduction}
The production of both charmonium mesons and beauty-flavoured hadrons,  
referred to as b-hadrons or \hb\ 
in this paper, in hadronic interactions represents a challenging 
testing ground for models based on Quantum ChromoDynamics (QCD).  

The mechanisms of \jpsi\ production operate at the boundary of the 
perturbative and non-perturbative regimes of QCD. At hadron colliders, 
\jpsi\ production was extensively studied at the Tevatron~\cite{Aco05,Abu07,Aba96,Abb99} 
and RHIC~\cite{Ada07}. 
Measurements in the new energy domain of the Large Hadron Collider (LHC) can contribute to a deeper 
understanding of the physics 
of the  
hadroproduction processes.  
The 
first LHC experimental results on the \jpsi\ transverse momentum (\pt) differential cross sections~\cite{Kha10,Aai11,Aad11,alice_jpsi,Chat12}  
are well described by various theoretical approaches~\cite{Lan09,But11,Ma11,Sal12}.  
Among those results, the ALICE Collaboration reported the measurement of the   
rapidity ($y$) and transverse momentum dependence of inclusive \jpsi\ production
in proton--proton (pp) collisions at $\sqrt{s}=7$~TeV~\cite{alice_jpsi}.
The inclusive \jpsi\ yield is composed of three contributions:
prompt \jpsi\ produced directly in the proton-proton collision,
prompt \jpsi\ produced indirectly (via the decay of heavier charmonium states 
such as \chic\ and ${\rm \psi(2S)}$),
and non-prompt \jpsi\ from the decay of b-hadrons.
Other LHC experiments have separated the prompt and non-prompt  \jpsi\  
component~\cite{Kha10,Aai11,Aad11,Chat12}. However, at mid-rapidity,    
only the high-\pt\ part of the differential $\dd \sigma_{\rm J/\psi} / \dd p_{\rm t}$ distribution 
was measured (\pt~$>6.5$~GeV/$c$), 
i.e. 
a 
small 
fraction (few percent) of the  
\pt-integrated cross section.   
%

The measurement of the production of b-hadrons in pp collisions at the LHC
provides a way to test, in a new energy domain,
calculations of QCD
processes based on the factorization approach.
In this scheme, the cross sections are computed as a convolution of the parton distribution
functions of the incoming protons, the partonic hard scattering cross sections,
and the fragmentation functions.
Measurements of cross sections for beauty quark production  
in high-energy hadronic interactions have been done in the past at ${\rm p \bar{p}}$ colliders at
center-of-mass energies from 630 GeV~\cite{UA1,Alb91} to 1.96 TeV~\cite{Abe95,Aba95,Abu07,Aco02} 
and in p-nucleus collisions with beam energies from 800 to 920 GeV~\cite{Hera-B}.  
The  LHC experiments have reported measurements of b-hadron production 
in pp collisions at $\sqrt{s}=7$~TeV by studying either exclusive decays of B mesons~\cite{CMSB+,CMSB0,CMSB0s} 
or semi-inclusive decays of b-hadrons~\cite{Kha10,Aai11,Aad11,Chat12,Aai10,Kha11}.   
At mid-rapidity, the measurements are available only for \pt\  of the b-hadrons larger 
than $\approx 5$  GeV/$c$, whereas the low \pt\ region of the differential b-hadron cross sections, 
where the bulk of the b-hadrons is produced, has not been studied.    

In this paper, the measurement of the fraction of \jpsi\ from the decay of b-hadrons 
in pp collisions at $\sqrt{s}=7$~TeV for \jpsi\ in the ranges  $1.3 < p_{\rm t} < 10$~GeV/$c$ and $|y|<0.9$ 
is determined.   
This information 
is combined with  
the previous inclusive \jpsi\ cross section measurement 
reported by ALICE~\cite{alice_jpsi}.  
Prompt \jpsi\ and b-hadron cross sections 
are thus determined at mid-rapidity down to the lowest \pt\ reach at the LHC energy.  

\section{Experiment and data analysis}
\label{due}
The ALICE experiment~\cite{Aam08} consists of a central barrel, covering the pseudorapidity region $|\eta|<$~0.9,   
and a muon spectrometer with $-4 < \eta < -2.5 $ coverage. The results presented in this paper  
were obtained with  
the central barrel tracking detectors, in particular 
the Inner Tracking System (ITS)~\cite{Aam08,alice_its} 
and the Time Projection Chamber (TPC)~\cite{alice_tpc}. 
The ITS, which  consists of two innermost Silicon Pixel Detector (SPD),
two Silicon Drift Detector (SDD), and two outer Silicon
Strip Detector (SSD) layers, provides up to six space points (hits) for each track.  
The TPC is a large cylindrical drift detector  
with an active volume that extends over the ranges
$85 < r < 247$~cm and $-250 < z < 250$~cm in the radial and longitudinal
(beam) directions, respectively. The TPC 
provides up to 159 
space points  
per track and 
charged particle identification via specific energy loss ($\dd E/\dd x$) measurement.  

The 
event sample, corresponding to  
$3.5 \times 10^8$ minimum bias events and an integrated luminosity $L_{\rm int} = 5.6\, {\rm nb^{-1}}$,  
event selection and track quality cuts used for the 
measurement of the inclusive \jpsi\ production at  
mid-rapidity~\cite{alice_jpsi} were also 
adopted in this analysis. In particular,   
an event with a reconstructed vertex position $z_{\rm v}$ was accepted if $|z_{\rm v}| < 10$~cm. 
The tracks were required to have a minimum \pt\ of 1~GeV/$c$, a minimum number of 70 TPC space points,  
a $\chi^2$ per space point of the
momentum fit lower than 4, 
and to point back to the interaction
vertex within 1 cm in the transverse plane. 
At least one hit in either of the two layers of the SPD was required. 
For tracks passing this selection, the average number of hits in the six ITS layers was 
4.5--4.7, depending on the data taking period.
The electron identification was based on 
the specific energy loss in the TPC: a $\pm 3 \sigma$ 
inclusion cut around the Bethe-Bloch fit for electrons and  $\pm 3.5 \sigma$ ($\pm 3 \sigma$) exclusion cut  
for pions (protons) were employed~\cite{alice_jpsi}. 
Finally, electron or positron candidates compatible, together with an opposite charge candidate,
with being products of $\gamma$ conversions   
(the invariant mass of the pair being smaller than 100 MeV/$c^2$) were removed, in
order to reduce the combinatorial background. It was verified, using
a Monte Carlo simulation, that this procedure does not affect the \jpsi\ signal.
In this analysis, opposite-sign (OS) electron pairs were divided in three 
``types'':  
type ``first-first'' ($FF$) 
corresponds to the case 
when  
both the electron and the positron have hits in the first pixel layer, 
type ``first-second'' ($FS$) 
are those pairs where 
one of them has a hit in the first layer and the other does not, 
while for the 
type ``second-second'' ($SS$) 
neither of them has a hit in the first layer. 
The candidates of type $SS$, which correspond to about 10\% of the total, were discarded due 
to the worse spatial resolution of the associated decay vertex.  

A detailed description of the track and vertex reconstruction procedures can be found in~\cite{alice_charm}.
The primary vertex was determined via an analytic $\chi^2$ 
minimization method 
in which tracks are approximated as straight lines 
after  propagation to their common point of closest approach.  
The vertex fit was constrained in the transverse plane using the information on the position and 
spread 
of the luminous region. The latter was  
determined from the distribution of primary vertices reconstructed over the run.  
Typically,  the transverse position of the vertex has a resolution that ranges from 
40~$\mu$m in low-multiplicity events with less than 10 charged particles per unit of rapidity   
to about 10~$\mu$m in events with a multiplicity of about 40. 
For each \jpsi\ candidate a specific primary vertex was also calculated 
by excluding the \jpsi\ decay tracks, in order to estimate a systematic 
uncertainty related to 
the evaluation of the primary vertex in the case of events with non-prompt \jpsi, as discussed in section~\ref{sec:syst}.   
The decay vertex of the \jpsi\ candidate was computed with the same analytic $\chi^2$ 
minimization as for the primary vertex, using the two decay tracks only and without the constraint of 
the luminous region.  

The measurement of the fraction of the \jpsi\ yield coming from b-hadron decays, $f_{\rm B}$,  relies on 
the discrimination of \jpsi\ mesons produced at a distance from the pp collision vertex. 
The signed projection of the \jpsi\ flight distance  
onto its transverse momentum vector, $\vec{p}_{\rm t}^{\rm J/\psi}$, 
was constructed according to the formula 
\begin{equation}
L_{xy}=\vec{L}\cdot \vec{p}_{\rm t}^{\rm J/\psi} / p_{\rm t}^{\rm J/\psi},  
\end{equation}
where $\vec{L}$ is the vector from the primary vertex to the \jpsi\ decay vertex.   
The variable $x$, referred to as ``pseudoproper decay length'' in the following, was introduced to separate 
prompt \jpsi\ from those produced by the decay of b-hadrons\footnote{
The variable $x$, which was introduced in~\cite{Aco05}, mimics a similar variable used for b-hadron lifetime
measurements  where b-hadrons are reconstructed exclusively and therefore the mass and \pt\ of
the b-hadron can be used in place of those of the \jpsi,  
to get $c\tau=\frac{L}{\beta\gamma}=\frac{c \cdot L_{xy}\cdot M_{\rm b-hadron}}{p_{\rm t}^{\rm b-hadron}}$.
},
\begin{equation}
x=\frac{ c \cdot L_{xy} \cdot m_{\rm J/\psi} }{p_{\rm t}^{\rm J/\psi}} , 
\label{eq:x}
\end{equation}
where $m_{\rm J/\psi}$ is the (world average) \jpsi\ mass~\cite{Nak10}.  

For events with very low \jpsi\ \pt , the non-negligible amount
of \jpsi\ with large opening angle between its flight direction and that of the b-hadron  
impairs the separation ability. Monte Carlo simulation shows that 
the detector resolution allows 
the determination of 
the fraction of \jpsi\ from the decay of b-hadrons   
for  events with \jpsi\ \pt\ greater than 1.3 GeV/$c$.  

An unbinned 2-dimensional 
likelihood fit was used 
to determine the ratio of the non-prompt to inclusive \jpsi\ production  
and the 
ratio of \jpsi\ signal candidates (the sum of both prompt and non-prompt components) to the 
total number of candidates, $f_{\rm Sig}$,  
by maximizing the quantity    
\begin{equation}
\ln L=  \displaystyle\sum\limits_{i=1}^N \ln F(x,m_{\rm e^+e^-}),   
\label{eq:logL}
\end{equation}
where $m_{\rm e^+e^-}$ is the invariant mass of the electron pair and 
$N$ is the total number of candidates in the range $2.4<m_{\rm e^+e^-}<4.0$~GeV/$c^2$.   
The expression for $F(x,  m_{\rm e^+e^-})$ is
\begin{equation}
F(x,  m_{\rm e^+e^-})=f_{\rm Sig}\cdot F_{\rm Sig}(x)\cdot M_{\rm Sig}(m_{\rm e^+e^-}) + 
                     (1-f_{\rm Sig}) \cdot F_{\rm Bkg}(x) \cdot M_{\rm Bkg}(m_{\rm e^+e^-}),  
\label{F}
\end{equation}
where 
$F_{\rm Sig}(x)$ and $F_{\rm Bkg}(x)$ are Probability Density Functions (PDFs) 
describing the pseudoproper decay length distribution for 
signal and background candidates, respectively. 
$M_{\rm Sig}(m_{\rm e^+e^-})$  and $M_{\rm Bkg}(m_{\rm e^+e^-})$ are the PDFs describing the dielectron invariant  
mass distributions for the signal and background, respectively.  
A Crystal Ball function~\cite{Gai82} is used for the  former and an exponential function for the latter.  
The 
signal PDF
is given by  
\begin{equation}
F_{\rm Sig}(x) = f_{\rm B}' \cdot F_{\rm B}(x) + (1-f_{\rm B}')\cdot F_{\rm prompt}(x),
\label{eq:Sig}
\end{equation}
where $F_{\rm prompt}(x)$ and $F_{\rm B}(x)$ are the PDFs 
for prompt and non-prompt \jpsi, 
respectively, and  $f_{\rm B}'$ is the fraction of reconstructed non-prompt \jpsi,   
\begin{equation}
f_{\rm B}'=\frac{N_{\rm J/\psi \leftarrow h_B}}{N_{\rm J/\psi \leftarrow h_B} + N_{\rm prompt\,  J/\psi}}, 
\end{equation}
which can differ (see below) from $f_{\rm B}$\ due to different acceptance and reconstruction efficiency of prompt 
and non-prompt \jpsi.  
The distribution of non-prompt \jpsi\ is the convolution of 
the $x$ distribution of \jpsi\ from b-hadron events, $\chi_{\rm B}(x)$, and 
the experimental resolution on $x$, $R_{type}(x)$, which depends on the type of candidate ($FF$ or $FS$),    
\begin{equation}
F_{\rm B}(x) =  \chi_{\rm B}(x') \otimes  R_{type}(x'-x).  
\end{equation}
Promptly produced \jpsi\ mesons decay at the primary vertex, and their pseudoproper decay length 
distribution is thus simply described  by $R_{type}(x)$:    
\begin{equation}
F_{\rm prompt}(x) = \delta(x') \otimes R_{type}(x'-x) = R_{type}(x).  
\end{equation}
The resolution function is described by the sum of two Gaussians and a  
power law function reflected about $x=0$ 
and was determined, as a function of the \pt\ of the \jpsi, 
with a Monte Carlo simulation study. In this simulation, which utilizes GEANT3~\cite{Bru94} and incorporates 
a detailed description of the detector material, geometry, and response,  
prompt \jpsi\ were generated   
with a \pt\ distribution extrapolated from CDF measurements~\cite{Aco05}  
and a $y$ distribution 
parameterization taken from  
Color Evaporation Model (CEM) calculations~\cite{Sto06}.    
These \jpsi\ were individually 
injected into 
proton--proton collisions simulated using the PYTHIA 6.4.21 event generator~\cite{Sjo94,Sjo06},    
and reconstructed as for \jpsi\ candidates in data.  
A data-driven method (discussed in section~\ref{sec:syst}) was also developed 
and used to estimate the systematic uncertainty related to this procedure.  
The Monte Carlo $x$ distribution of \jpsi\ from the decay of b-hadrons produced in proton-proton collisions
simulated using the PYTHIA 6.4.21 event generator~\cite{Sjo94,Sjo06} with Perugia-0 tuning~\cite{Ska10}
was taken as the template for the $x$ distribution of b-hadron events in data, $\chi_{\rm B}(x)$.   
A second template, used to estimate the systematic uncertainty, was obtained by decaying the 
simulated b-hadrons using the EvtGen package~\cite{evtgen},  
and describing the final state radiation (``internal'' bremsstrahlung) using PHOTOS~\cite{photos1,photos2}. 


For the background $x$ distribution, $F_{\rm Bkg}(x)$, the functional form employed by 
CDF~\cite{Aco05} was used,  
\begin{equation}
\begin{split}
F_{\rm Bkg}(x)=& (1- f_+ - f_- - f_{\rm sym}) R_{type}(x)  \\
               + & \left[ 
               \frac{f_+}{\lambda_+} e^{-x'/\lambda_+}\theta(x') +
               \frac{f_-}{\lambda_-} e^{x'/\lambda_-}\theta(-x') + 
               \frac{f_{\rm sym}}{2\lambda_{\rm sym}} e^{-|x'|/\lambda_{\rm sym}} 
               \right] \otimes  R_{type}(x'-x), 
\end{split}
\label{eq:BkgX}
\end{equation}
where $\theta(x)$ is the step function, $f_+$, $f_-$ and $f_{\rm sym}$ are the fractions of three 
components with positive, negative and symmetric decay length exponential distributions, respectively. 
The effective parameters $\lambda_+$, $\lambda_-$ and $\lambda_{\rm sym}$, and optionally also  
the corresponding fractions, were determined, prior to  the likelihood fit maximization, with a 
fit to the $x$\ distribution in the  sidebands of the dielectron invariant mass distribution, 
defined as the regions 1.8--2.6 and 3.2--5.0~GeV/$c^2$. 
The introduction of these components is needed because 
the background consists also of random combinations of electrons from semi-leptonic 
decays of charm and beauty hadrons, which tend to produce positive $x$ values, 
as well as of other secondary or mis-reconstructed tracks which contribute both to positive and 
negative $x$ values. The first term in eq.~\ref{eq:BkgX}, proportional to $R_{type}(x)$, 
describes the residual combinatorics of primary particles.  

\begin{figure}[tb]
\centering
\includegraphics[width=.49\textwidth]{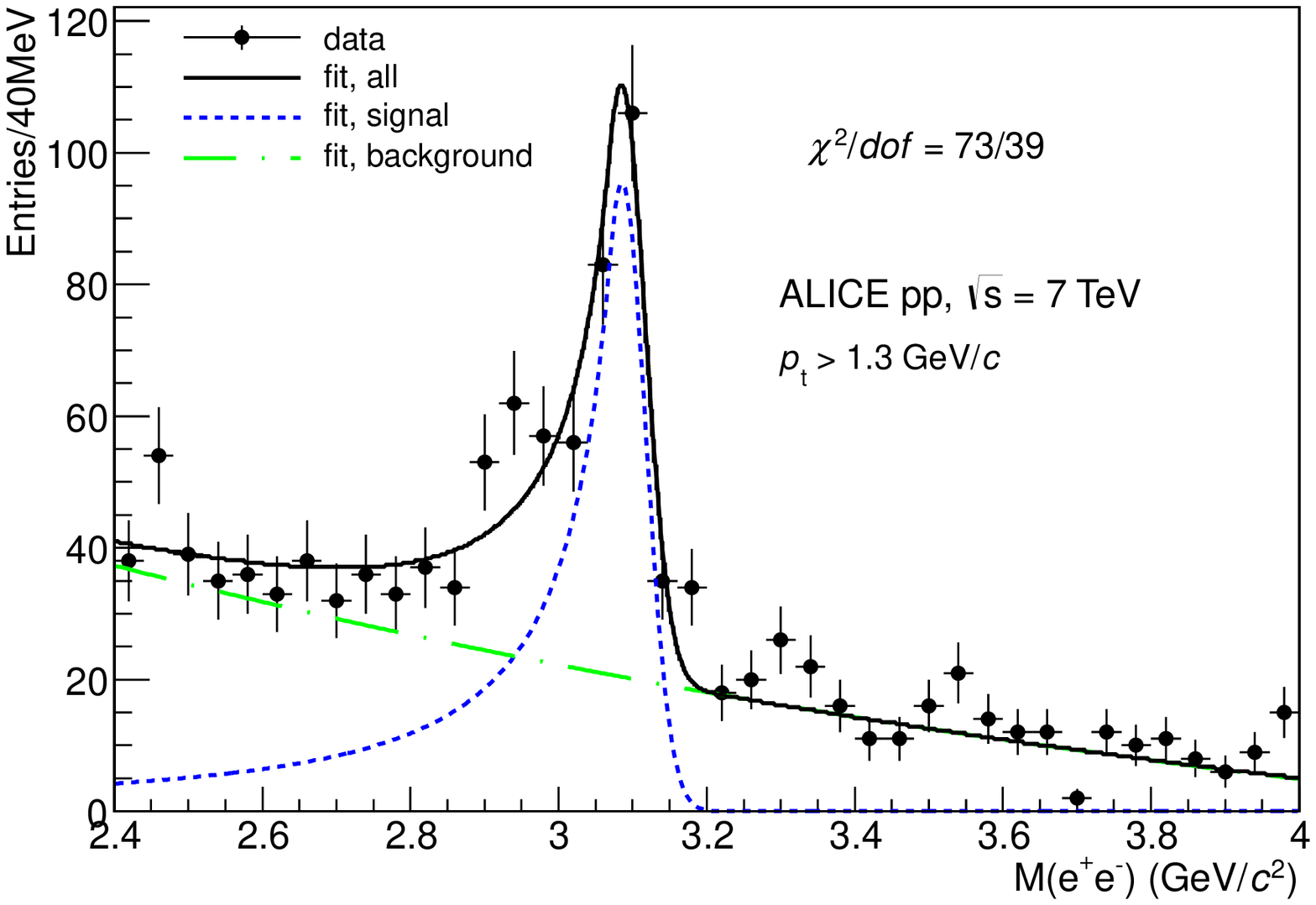}
\includegraphics[width=.49\textwidth]{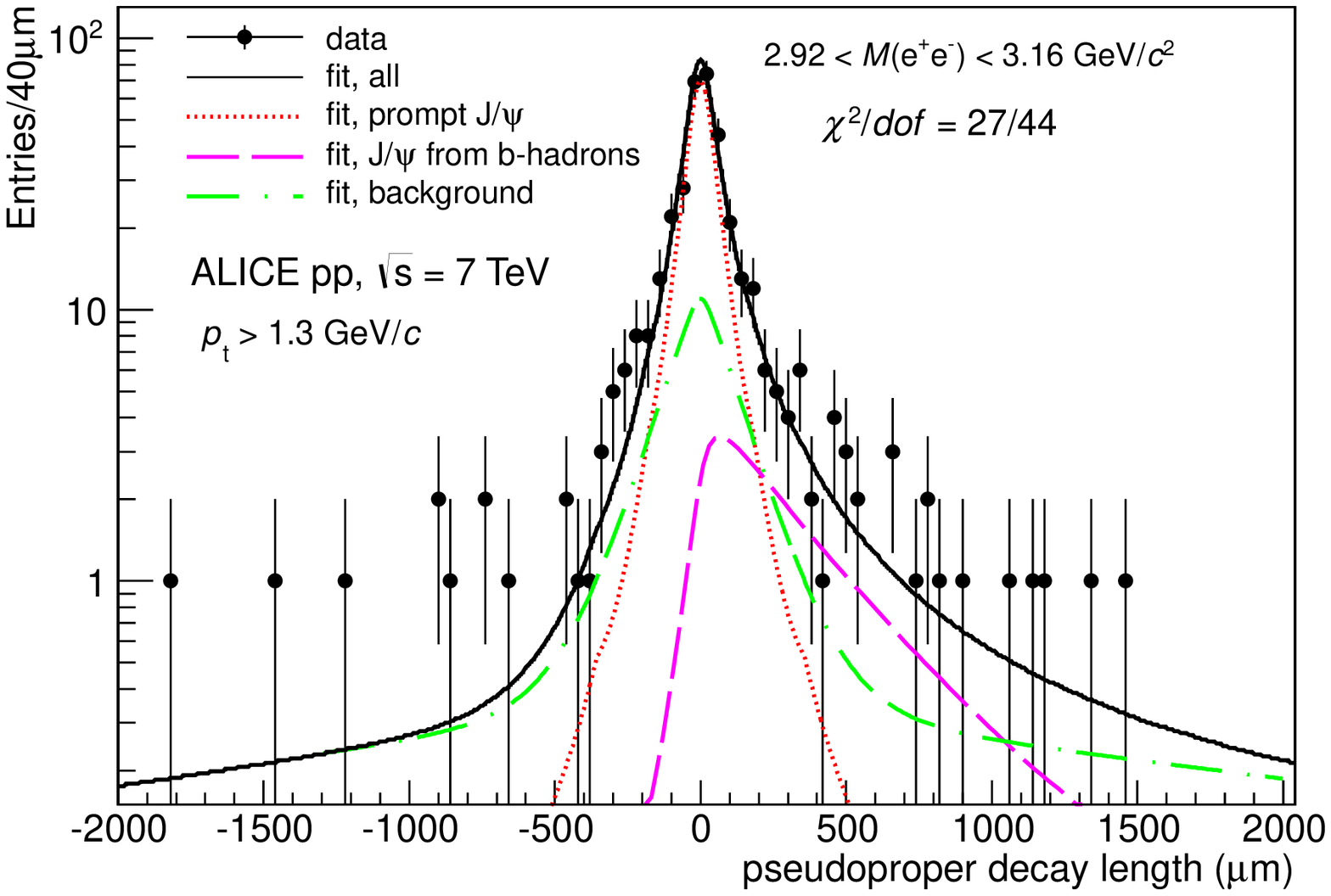}
\caption{Invariant mass (left panel) and pseudoproper decay length (right panel)  
         distributions of opposite sign electron pairs  
         for $|y_{\rm J/\psi}|<0.9$\ 
         and $p_{\rm t}^{\rm J/\psi} > 1.3$~GeV/$c$      
         with superimposed projections of the maximum likelihood fit.   
         The latter distribution is limited to the \jpsi\ candidates under the mass peak, 
         i.e. for $2.92 < m_{\rm e^+ e^-} < 3.16$~GeV/$c^2$, for 
         display purposes only. 
         The $\chi^2$ values of these projections are reported for both distributions.  
}
\label{fig:1}
\end{figure}

In figure~\ref{fig:1} the distributions of 
the invariant mass 
and the pseudoproper decay length, the latter restricted to  candidates
with $2.92 < m_{\rm e^+ e^-} < 3.16$~GeV/$c^2$, 
for opposite-sign electron pairs with \pt~$>1.3$~GeV/$c$ 
are shown with superimposed projections of the maximum likelihood fit result.   

The value of the fit parameter $f_{\rm B}'$\ provides the fraction of non-prompt \jpsi\ which 
were  
reconstructed. In principle prompt and non-prompt \jpsi\ can have different  
acceptance times efficiency ($A \times \epsilon$) values. This can happen because of two effects: 
{\em (i)}
the $A \times \epsilon$\ depends on the \pt\ of the \jpsi\ and prompt and non-prompt \jpsi\ have 
different \pt\ distributions within the considered \pt\ range; 
{\em (ii)} at a given \pt, prompt and non-prompt \jpsi\ can have different polarization and, 
therefore, a different acceptance.  
The fraction of non-prompt \jpsi, corrected for these effects, was obtained as  
\begin{equation}
f_{\rm B} = \left( 1 + \frac{1-f_{\rm B}'}{f_{\rm B}'} \cdot 
 \frac{\langle A \times \epsilon \rangle_{\rm B}}{\langle A \times \epsilon \rangle_{\rm prompt}} \right)^{-1},  
\label{correction}
\end{equation}
where $\langle A \times \epsilon \rangle_{\rm B}$ and $\langle A \times \epsilon \rangle_{\rm prompt}$
are the average acceptance times efficiency values, in the considered \pt\ range and for the 
assumed polarization state, of non-prompt and prompt \jpsi, respectively. 
The acceptance times efficiency ($A\times \epsilon$) varies very smoothly with \pt\  
and, for unpolarized \jpsi\ in the \pt\ range from 1.3 to 10 GeV/$c$, has a minimum 
of 8\% at 2 GeV/$c$ and a broad maximum of 12\% at 7 GeV/$c$~\cite{alice_jpsi}.  
As a consequence, the $\langle A\times \epsilon \rangle$ values of prompt
and non-prompt \jpsi\ differ  by about 3\% only in this integrated \pt\ range.     


The central values of the 
resulting cross sections  
are quoted  assuming both 
prompt and non-prompt \jpsi\ to be unpolarized and the variations due to different assumptions 
are estimated as a separate systematic uncertainty.  
The polarization of \jpsi\ from  b-hadron  
decays is expected to be much smaller than for prompt \jpsi\ due to the averaging
effect caused by the admixture of various exclusive ${\rm B\rightarrow J/\psi +X}$  decay channels.  
In fact, the sizeable polarization, which is  observed when the polarization axis 
refers to the B-meson direction~\cite{Aub03}, is strongly smeared when 
calculated with respect to the direction of the daughter \jpsi~\cite{Aai11}, 
as indeed observed by CDF~\cite{Abu07}. 
Therefore, these variations will be calculated in the two cases of prompt \jpsi\ with 
fully transverse ($\lambda = 1$) or longitudinal ($\lambda=-1$) 
polarization, in the Collins-Soper (CS)  
and helicity (HE) reference frames\footnote{The polar angle distribution of the \jpsi\ decay leptons is given by
$\dd N/\dd \cos \theta = 1+ \lambda \cos ^2 \theta$.}, the non-prompt component being left unpolarized.

\begin{figure}[tb]
\centering
\includegraphics[width=.49\textwidth]{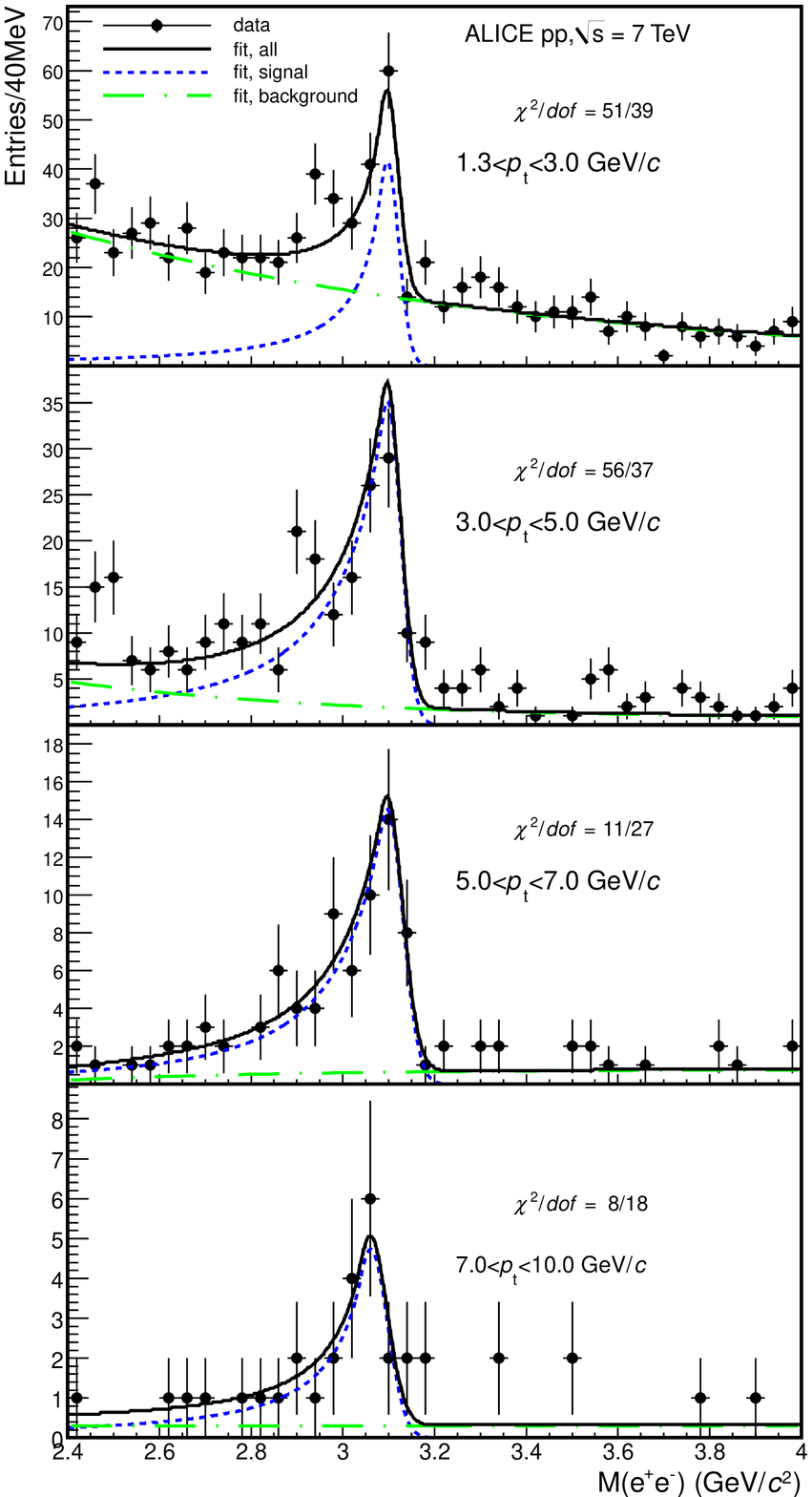}
\includegraphics[width=.49\textwidth]{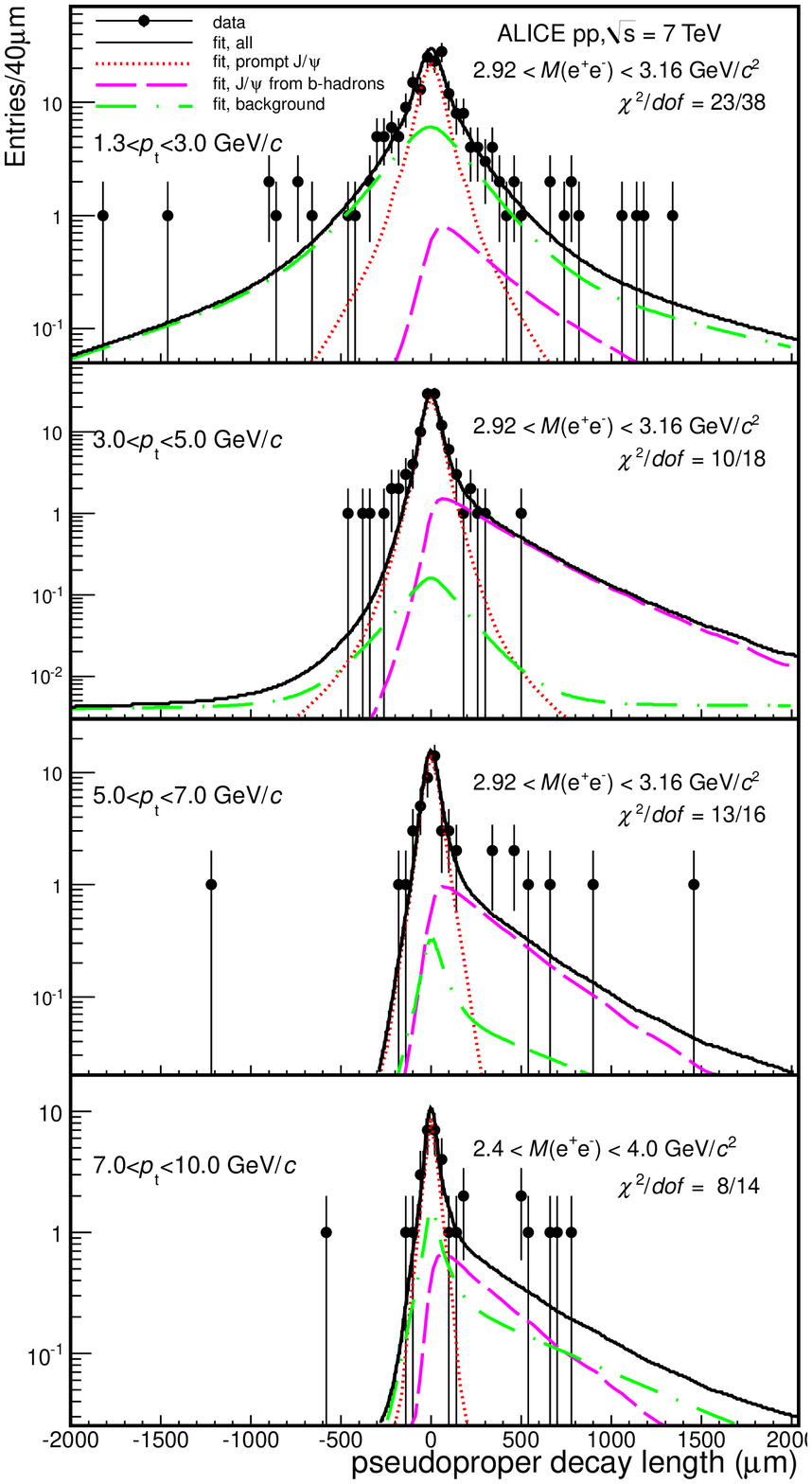}
\caption{Invariant mass (left panels) and  pseudoproper decay length (right panels)
         distributions in different \pt\ bins  
         with superimposed  projections  of the maximum likelihood fit. 
         The $\chi^2$ values of these projections are also reported for all distributions. 
}
\label{fig:2}
\end{figure}

%
%
Despite the small \jpsi\ candidate yield, amounting to about 400  
counts, 
the data sample could be divided into four \pt\ bins (1.3--3,  3--5, 5--7 and 7--10 GeV/$c$), 
and the fraction $f_{\rm B}$ was  evaluated in each of them with the same technique. 
At low \pt\ the statistics is higher, but the resolution is worse and  
the signal over background, $S/B$, is smaller (i.e. $f_{\rm Sig}$ is smaller).   
At high \pt\ the statistics is smaller, but the resolution improves and 
the background becomes negligible. 
In figure~\ref{fig:2} the distributions of the invariant mass and of the pseudoproper
decay length are shown in different \pt\ bins with superimposed results of the fits.

\section{Systematic uncertainties}
\label{sec:syst}
The different contributions to the systematic uncertainties affecting the  measurement
of the fraction of \jpsi\ from the decay of b-hadrons are discussed in the following, 
referring to the integrated \pt\ range,  
and summarized in table~\ref{tab:1}.

\begin{itemize}

\item {\bf Resolution function.}
The resolution function was determined from a Monte Carlo simulation, as discussed above. 
The fits were repeated by artificially modifying  the resolution function, according to the formula
\[
R_{type}'(x)=\frac{1}{1+\delta} R_{type}\left( \frac{x}{1 + \delta} \right), 
\]  
where $\delta$ is a constant representing the desired relative variation of the RMS of the resolution 
function.   
Studies on track distance of closest approach to the primary interaction vertex 
in the bending plane ($d_0$) 
show that the \pt\ dependence of the $d_0$ resolution as measured in the data  
  is reproduced 
  within about 10\% by the Monte Carlo simulation~\cite{alice_charm}, 
  but with a systematically worse resolution in data.  
For the $x$ variable  a similar direct comparison to data is not straightforward, 
however, the residual discrepancy is not expected to be larger than that observed for $d_0$.

The variations of $f_{\rm B}$ obtained in the likelihood fit results by varying $\delta$
from $-5\%$ to $+10\%$
are +8\% and --15\%, respectively, and
they were 
assumed as the systematic uncertainty due to this contribution.  

An alternative, data-driven,  approach was also considered. 
The $x$ distribution of the signal, composed of prompt and non-prompt \jpsi, was obtained by 
subtracting the $x$ distribution of the background, 
measured in the sidebands of the invariant mass distribution. This distribution is then fitted by fixing 
the ratio of prompt to non-prompt \jpsi\ to that obtained from the likelihood fit 
and leaving free the parameters of the resolution function. The RMS of the fitted resolution 
function is found to be 
8\% larger than the one  
determined using the Monte Carlo simulation, 
hence within the range of variation assumed for $\delta$.  


\item {\bf Pseudoproper decay length distribution of background.}
The shape of the combinatorial background was determined from a fit to the $x$ distribution of candidates in 
the sidebands of the invariant mass distribution. By varying the fit parameters within their 
errors an envelope of distributions was obtained, whose extremes were used in the likelihood fit in place 
of the most probable distribution. The variations in the result of the fit were 
determined and adopted as systematic uncertainties.  
Also, it was verified that the $x$\ 
distribution obtained for like-sign (LS) candidates, with invariant mass 
in the range from 2.92 to 3.16 GeV/$c^2$ complementary to the sidebands,    
is best fitted by a distribution which falls within the envelope of the OS 
distributions. Finally, the likelihood fit was repeated by relaxing, 
one at a time, 
the parameters of the functional form (eq.~\ref{eq:BkgX}) and 
it was found  
that  
the values of $f_{\rm B}$ were within the estimated uncertainties. 
The estimated systematic uncertainty is $ 6\%$.   

\item{\bf Pseudoproper decay length distribution of b-hadrons.} 
The fits were also done using as template for the $x$ distribution of b-hadrons, $\chi_{\rm B}(x)$, that 
obtained by the EvtGen package~\cite{evtgen},
and describing the final state radiation using PHOTOS~\cite{photos1,photos2}. 
The central values of the fits differ by a few percent at most and the resulting systematic uncertainty is $3\%$.    
\item {\bf Invariant mass distributions.}  
The likelihood method was used in this analysis to fit simultaneously the invariant mass distribution,  
which is sensitive to the ratio of signal to all candidates ($f_{\rm Sig}$), 
and the $x$ distribution, which determines the ratio of non-prompt to signal candidates ($f_{\rm B}$). 
The statistical uncertainties on these quantities were therefore evaluated together,  
including the effects of correlations. However, the choice of the function describing the 
invariant mass distribution, as well as the procedure, can introduce  
systematic uncertainties in the evaluation of $f_{\rm B}$.  
Different approaches were therefore considered:  
{\em (i)} the functional form describing the background was changed 
          into an exponential plus a constant and the fit repeated; 
{\em (ii)} the background was described using the LS distribution and the signal was obtained by 
           subtracting the LS  from the OS distributions. 
           The signal and the background shapes were determined with 
           $\chi^2$ minimizations.  
           Both functional forms, exponential and exponential plus a constant, 
           were considered for the background.  
           The likelihood fit was then performed
           again to determine $f_{\rm B}$ (and $f_{\rm Sig}$); 
{\em (iii)} the same procedure as in  {\em (ii)} was used, but additionally 
            $f_{\rm Sig}$ was estimated {\em a priori} using a bin counting 
            method~\cite{alice_jpsi} instead of the integrals of the best fit functions.  
            The maximum likelihood fit was performed with $f_{\rm Sig}$ fixed to this new value;  
{\em (iv)} and {\em (v)}  the same procedures as in {\em (ii)} and {\em (iii)} were used but 
           with the background described by a track rotation (TR) method~\cite{alice_jpsi}.  

%
%
%
Half of the difference between the maximum and minimum $f_{\rm B}$ values obtained with the different methods   
was assumed as systematic uncertainty. It amounts to about $6\%$.
\item{\bf Primary vertex.} The effect of excluding the decay tracks of the \jpsi\ candidate in 
the computation of the primary vertex was studied with the Monte Carlo simulation:    
on the one hand, for the prompt \jpsi, the $x$ resolution function is degraded, due to the 
fact that two prompt tracks are not used in the computation of the vertex, which is thus determined 
with less accuracy. The effect on the resolution is 
\pt\ dependent, 
with the RMS of the $x$ distribution of prompt \jpsi\ increasing by 15\% at low \pt\ and by 7\% at high \pt.
On the other hand, for non-prompt \jpsi\ a bias on the $x$\ determination 
should be reduced.  
The bias consists in an average shift of the primary vertex towards the secondary decay vertex of  
the b-hadrons, which is reflected in a shift of the mean of the $x$ distribution by about $4$~$\mu$m 
for the \pt-integrated distribution. However, the shift is \pt\ and ``type'' dependent.     
In some cases the bias is observed in the opposite direction  
and is enhanced by removing the decay tracks of the candidate.   
This can happen since b-quarks are always produced in pairs. If a charged track from the fragmentation of 
the second b-quark also enters the acceptance, it can pull the primary vertex position towards the opposite direction.
In the end, therefore, the primary vertex was computed without removing the decay 
tracks of the candidates.   
To estimate the systematic uncertainty, the analysis was repeated by either 
{\em (i)} removing the decay tracks in the computation of the primary vertex and using the corresponding 
          worse resolution function in the fit or 
{\rm (ii)} keeping those tracks and introducing an {\em ad hoc} shift in the distribution of the 
$\chi_{\rm B}(x)$, equal to that observed in the Monte Carlo simulation for non-prompt \jpsi.   
The contribution to the systematic  uncertainty is about $5\%$.   

\item {\bf MC $p_{\rm t}$\ spectrum.} The ratio  
$\frac{\langle A \times \epsilon \rangle_{\rm B}}{\langle A \times \epsilon \rangle_{\rm prompt}}$
in eq.~\ref{correction} 
was computed using  MC simulations: prompt \jpsi\ were generated with the \pt\ distribution 
extrapolated from CDF measurements~\cite{Aco05} and the $y$ distribution parameterized from CEM~\cite{Sto06}; 
b-hadrons were generated using the PYTHIA 6.4.21~\cite{Sjo94,Sjo06} event generator with 
Perugia-0 tuning~\cite{Ska10}.  
By varying the 
average \pt\ 
of the \jpsi\ distributions 
within a factor 2, a 1.5\% variation in the acceptance was obtained both for prompt and non-prompt \jpsi. 
Such a small value is a consequence of the weak \pt\ dependence of the acceptance.  
For the measurement integrated over \pt\ (\pt$>~1.3$~GeV/$c$), the $A \times \epsilon$\ values of prompt and non-prompt \jpsi\ 
differ by about 3\% only. The uncertainty due to Monte Carlo \pt\ distributions is thus estimated to be 1\%.   
When estimating $f_{\rm B}$ in \pt\ bins, this uncertainty is negligible.  
\item {\bf Polarization.} The variations of $f_{\rm B}$ obtained assuming  different 
polarization scenarios for the prompt component only were evaluated, as discussed in section~\ref{due}, and are reported 
in table~\ref{tab:1}.   
The maximum variations are quoted as separate errors.  
\end{itemize}

\begin{table}
\caption{\label{tab:1}
Systematic uncertainties (in percent) on the measurement of the 
fraction of \jpsi\ from the decay of b-hadrons, $f_{\rm B}$.    
The variations of $f_{\rm B}$ are also 
reported, with respect to the case of both prompt and non-prompt \jpsi\ unpolarized, when assuming 
the prompt component with given polarization.    
}
\centering
\begin{tabular}{l| c |c  |c}
\hline
{\bf Source} & \multicolumn{3}{c} {\bf Systematic uncertainty (\%)}  \\
       &  \pt\ integrated  &   lowest \pt\ bin &   highest \pt\ bin\\  
\hline
Resolution function & +8, --15 & +15, --25  & +2,  --3 \\
$x$ distribution of background & $\pm 6$ & $\pm 13$  &  $\pm 1$   \\
$x$ distribution of b-hadrons & $\pm 3$  & $\pm 3$ & $\pm 2$ \\
$m_{\rm e^+e^-}$ distributions & $\pm 6$ & $\pm11$ &  $\pm4$ \\
Primary vertex   & +4, --5  & $\pm 4$  & +4, --8 \\
MC \pt\ spectrum & $\pm 1$  & 0 & 0\\
Total & +12, --18  & +23, --30 & +6, --9 \\
\hline
Polarization (prompt \jpsi) &   &  & \\ \hline
\multicolumn{1}{l|}{ \; CS ($\lambda=-1$)} & +13 & +22&  +5  \\ 
\multicolumn{1}{l|}{ \; CS ($\lambda=+1$)} &--10 &--19& --3  \\
\multicolumn{1}{l|}{ \; HE ($\lambda=-1$)} & +17 & +19& +11  \\    
\multicolumn{1}{l|}{ \; HE ($\lambda=+1$)} &--14 &--16& --8  \\ \hline
\end{tabular}
\end{table}

The study of systematic uncertainties was repeated as a function of \pt. In table~\ref{tab:1} the 
results are summarized for the integrated \pt\ range (\pt~$>1.3$~GeV/$c$) and for the lowest (1.3--3~GeV/$c$) 
and highest (7--10~GeV/$c$) \pt\ bins. All systematic uncertainties increase with decreasing \pt, 
except the one related to the primary vertex measurement.    

\section {Results}
\subsection{Fraction of J/$\psi$ from the decay of b-hadrons}
\begin{figure}[tb]
\centering{\includegraphics[width=.55\textwidth]{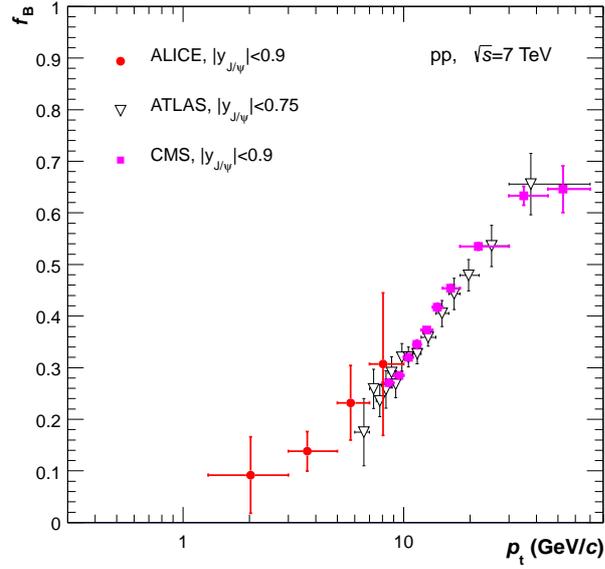}} 
\caption{The fraction of \jpsi\ from the decay of b-hadrons as a function 
of \pt\ of \jpsi\ compared with results 
         from ATLAS~\cite{Aad11} and CMS~\cite{Chat12} in pp collisions at $\sqrt{s}$~=7~TeV.   
}
\label{fig:3}
\end{figure}

The fraction of \jpsi\ from the decay of b-hadrons in the experimentally accessible kinematic range,  $p_{\rm t}>1.3$~GeV/$c$ and $|y|<0.9$,  
which is referred to as ``measured region'' in the following,   
is 
\[
f_{\rm B}=0.149 \pm 0.037\,{\rm (stat.)} ^{+0.018}_{-0.027}\, {\rm (syst.)} ^{+0.025 \, (\lambda_{\rm HE}=1)}_{-0.021\, (\lambda_{\rm HE}=-1)}\,{\rm (syst. pol.)}. 
\]
The fractions measured in the \pt\ bins are reported in table~\ref{tab:2} and shown in 
figure~\ref{fig:3}. In the figure, the data symbols are placed at the average value of the \pt\ distribution of each bin. 
The average was computed using the above mentioned Monte Carlo distributions: the one based on the CDF extrapolation~\cite{Sto06}  
and  that using PYTHIA~\cite{Sjo94,Sjo06} with Perugia-0 tuning~\cite{Ska10} 
for prompt and non-prompt \jpsi, respectively, weighted by the measured $f_{\rm B}$.  
In figure~\ref{fig:3} the results of the 
ATLAS~\cite{Aad11}  
and CMS~\cite{Chat12}
experiments measured at mid-rapidity for the same colliding system 
 are also shown.  
The ALICE results extend the mid-rapidity measurements 
down to low \pt.

\begin{table}[t]
\caption{\label{tab:2}
The   
fraction of \jpsi\ from the decay of b-hadrons and cross sections.   
Some of the contributions to the systematic uncertainty do not depend on \pt,
thus affecting only the overall normalization, and they are separately
quoted (correl.). The contributions which depend on \pt, 
even when they are correlated bin by bin, were included among
the non-correlated systematic errors. The values of $\langle p_{\rm t}\rangle$  were 
computed using Monte Carlo distributions (see text for details).    
}
{\footnotesize
\centering
\begin{tabular}{l l  l  ll l ll }
\hline\hline
 \pt\    & $\langle p_{\rm t}\rangle$ & Measured & \multicolumn{5}{l}{Systematic uncertainties} \\  \cline{4-8} 
$\left(  {\rm GeV}/c \right)$ &  $\left(  {\rm GeV}/c \right)$   & quantity & Correl. & Non-correl. & Extrap. & Polariz., CS & Polariz., HE   \\
\hline\hline 
%
%
            &   &  $ f_{\rm B}$ (\%)  &  & & & &  \\ 
 1.3--3.0   & 2.02  &$9.2\pm7.4$ &0&+2.1, --2.8& 0 & +2.0, --1.7  & +1.7, --1.5 \\
 3.0--5.0   & 3.65  &$13.8\pm3.8$ &0&+1.5, --2.1& 0 & +1.3, --1.0  & +2.1, --3.0 \\
 5.0--7.0   & 5.75  &$23.2\pm7.2$ &0&+1.6, --2.1& 0 & +0.2, --0.2  & +3.5, --2.6 \\
 7.0--10.0  & 8.06  &$30.7\pm13.8$ &0&+1.8, --2.8& 0 & +1.5, --0.9  & +3.4, --2.5 \\ 
 \pt~$>1.3$  & 2.85  &$14.9\pm3.7$ &0&+1.8, --2.7&     0     & +1.9, --1.5  & +2.5, --2.1  \\
 \pt~$>0$    & 2.41  &$14.3\pm3.6$ &0&+1.8, --2.6&+0.2, --0.5& +2.4, --1.6  & +2.5, --1.9  \\ \hline   
            &   &\multicolumn{3}{l}{{\bf ${\rm d}^2 \sigma_{\rm J/\psi}/{\rm d}y {\rm d}p_{\rm t}$} \; $\left( \frac{\rm  nb}{{\rm GeV}/c}\right)$} \\
 1.3--3.0   & 2.02 &$1780\pm 210$ & $\pm65$ & $\pm250$ & 0  & +400, --320   & +330, --280\\ 
 3.0--5.0   & 3.65 &$715\pm125$   & $\pm25$ & $\pm90$ & 0  &+50, --60       & +170, --90 \\
 5.0--7.0   & 5.74 &$405\pm70$    & $\pm15$ & $\pm45$ & 0  & +1, --3       & +50, --50  \\
 7.0--10.0  & 8.06 &$60\pm25$     & $\pm2$ & $\pm12$ & 0  & +2, --3        & +5, --6  \\ \hline
%
            &   &\multicolumn{3}{l}{{\bf ${\rm d}^2 \sigma_{\rm prompt \; J/\psi}/{\rm d}y {\rm d}p_{\rm t}$} \; $\left( \frac{\rm nb}{{\rm GeV}/c}\right)$}  \\
 1.3--3.0   & 2.02 &$1600\pm 230$ & $\pm60$    &$\pm230$   & 0 & +400, --320 & +330, --280 \\
 3.0--5.0   & 3.65 &$620\pm 110$& $\pm20$      & $\pm 80$ & 0 &+50, --60 & +170, --90 \\
 5.0--7.0   & 5.74 &$310\pm60$& $\pm 10$     & $\pm 35$ & 0 &+1,   --3 & +50,  --50 \\
 7.0--10.0  & 8.03 &$40\pm18$ & $\pm 1$      & $\pm 8$  & 0 &+2,   --3 & +5,    --6 \\ \hline \smallskip
%
%
           &   &\multicolumn{3}{l}{$\sigma_{\rm prompt \; J/\psi}(|y_{\rm J/\psi}|< 0.9) $ \; ($\mu$b)} \\
 \pt~$>$1.3& 2.81 & $8.3\pm0.8$ & \multicolumn{2}{c}{$\pm 1.1$}  &    0         & +1.0, --1.2 &  +1.5, --1.4  \\
 \pt~$>$0 & 2.37 & $10.6\pm1.1$ & \multicolumn{2}{c}{$\pm 1.6$ } & +0.06, --0.02& +1.6, --1.7 & +1.9, --1.8   \\ \hline
%
%
           &   &\multicolumn{3}{l}{$\sigma_{{\rm J/\psi} \leftarrow {\rm h_B}}(|y_{\rm J/\psi}|< 0.9) $ \; ($\mu$b)} \\
 \pt~$>$1.3 & 3.07 & $1.46\pm0.38$& \multicolumn{2}{c}{+0.26, --0.32}&    0    &    0     &    0     \\
 \pt~$>$0   & 2.62 & $1.77\pm0.46$& \multicolumn{2}{c}{+0.32, --0.39}& +0.02, --0.06&   0 &  0        \\ \hline
%
%
           &   &\multicolumn{3}{l}{ $ {\rm d} \sigma_{\rm b\bar{b}}/{\rm d}y\big|_{|y|<0.9} $ \; ($\mu$b) } \\
           &   &$43\pm 11$&    \multicolumn{2}{c}{+9, --10} &+0.6, --1.5 & 0 & 0         \\ \hline
           &   &\multicolumn{3}{l}{{\bf $\sigma_{\rm b\bar{b}}$} \; ($\mu$b) } \\
           &   &$282\pm 74$&      \multicolumn{2}{c}{+58, --68} & +8, --7 &  0 &  0 \\
\end{tabular}
}
\end{table}
\subsection{Prompt J/$\psi$ production}
By combining the measurement of the inclusive \jpsi\ cross section, which was determined as described in~\cite{alice_jpsi}, and the $f_{\rm B}$ value,  
the prompt \jpsi\ cross section was obtained: 
\begin{equation} 
\sigma_{\rm prompt \; J/\psi} = (1-f_{\rm B}) \cdot \sigma_{J/\psi}. 
\end{equation}
The numerical values of the inclusive \jpsi\ cross section in the \pt\ ranges used for this analysis  
are summarized in table~\ref{tab:2}. 
In the measured region 
the integrated cross section is 
$
\sigma_{\rm prompt \; J/\psi}(|y|<0.9, p_{\rm t}>1.3\, {\rm GeV/}c) = 
8.3 \pm 0.8 {\rm (stat.)} \pm 1.1 {\rm (syst.)} ^{+1.5 {\rm (\lambda_{HE}=1)}}_{-1.4 {\rm (\lambda_{HE}=-1)}} \, {\rm \mu b}. 
$
The systematic uncertainties related to the unknown polarization
are quoted for the reference frame where they are the largest. 

\begin{figure}[tb]
\centering{\includegraphics[width=.48\textwidth]{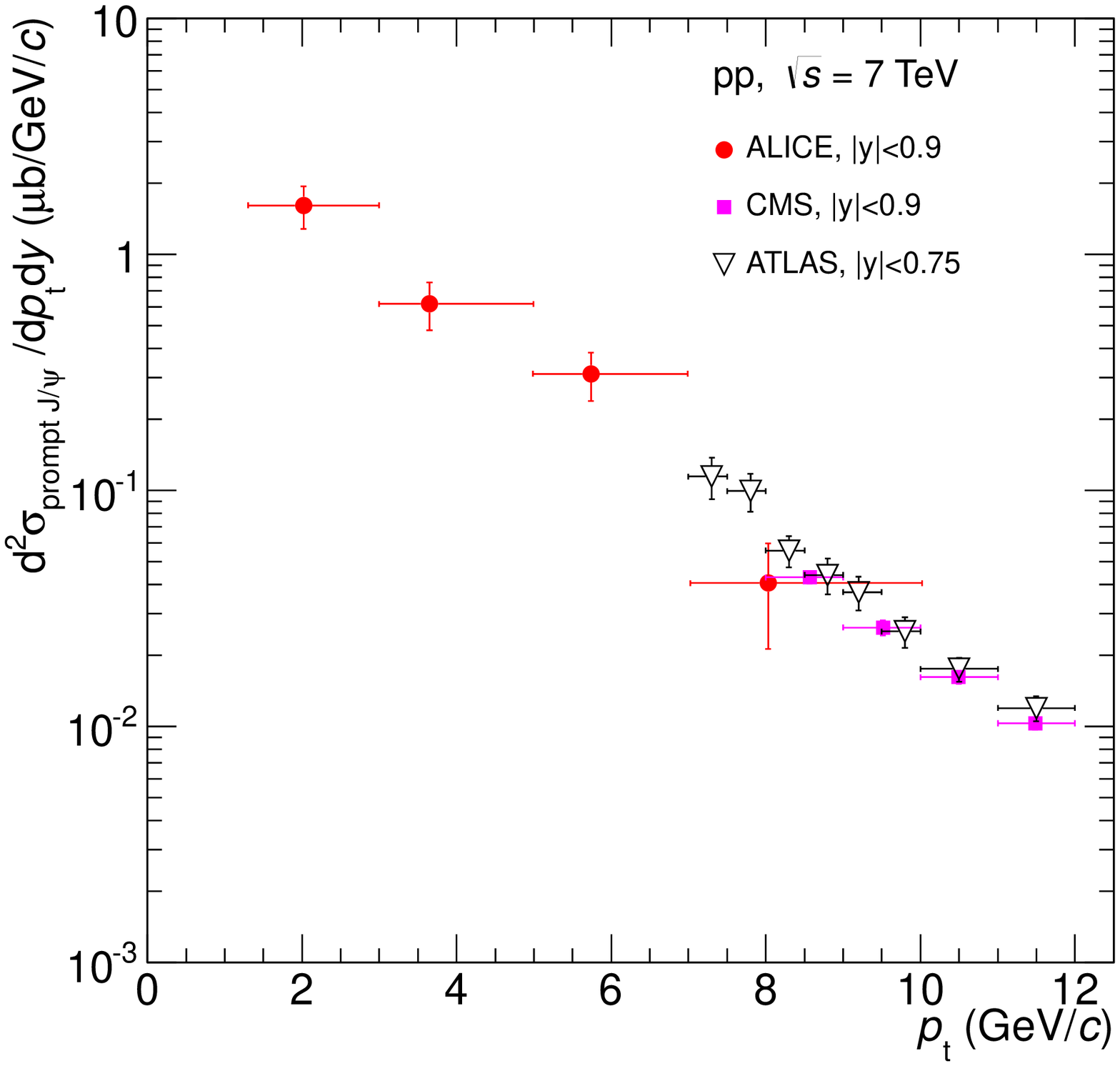}}
\centering{\includegraphics[width=.48\textwidth]{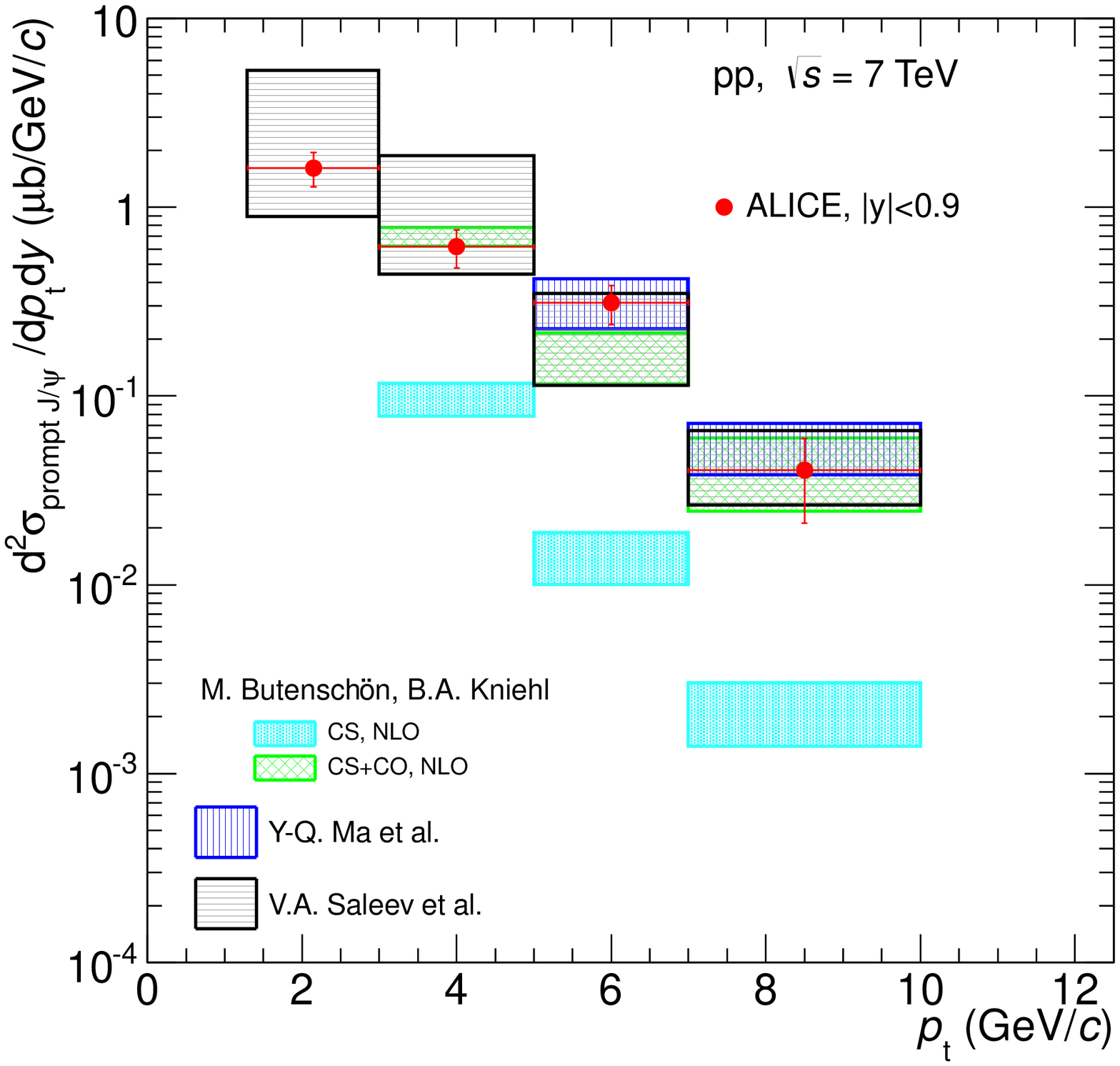}}
\caption{$\frac{\dd \sigma_{\rm prompt \; J/\psi}}{\dd p_{\rm t}\dd y}$\ as a function of \pt\ 
compared to results from 
ATLAS~\cite{Aad11} and CMS~\cite{Chat12} 
at 
mid-rapidity (left panel) 
and to 
theoretical 
calculations~\cite{Ma11,But11,Sal12} (right panel).  
The error bars represent the quadratic sum of the statistical and
systematic uncertainties.  
}
\label{fig:4}
\end{figure}

\begin{figure}[tb]
\centering{\includegraphics[width=.60\textwidth]{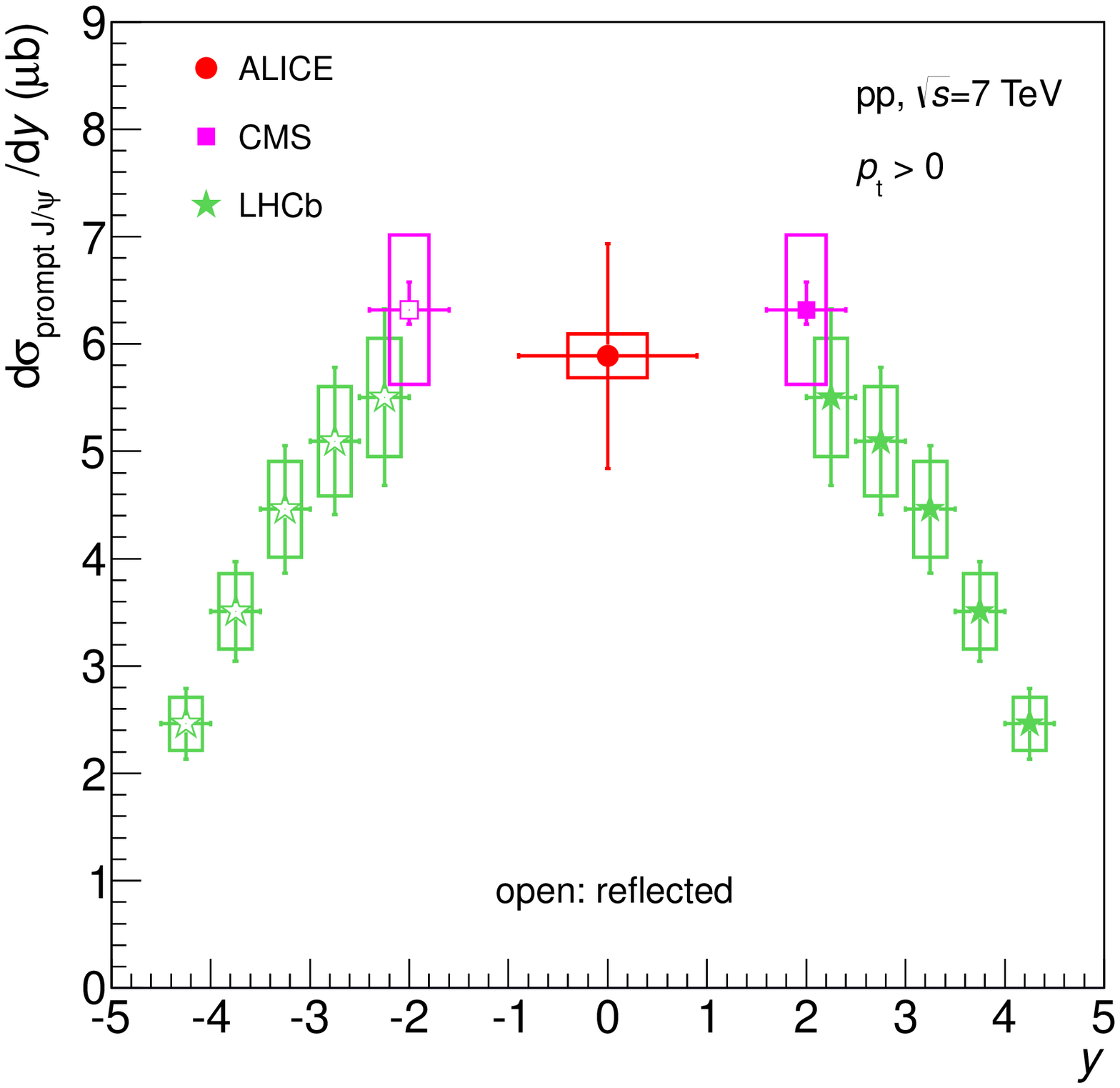}} 
\caption{$\frac{\dd  \sigma_{\rm prompt \; J/\psi}}{\dd y}$\ as a function of $y$. 
The error bars represent the quadratic sum of the statistical and systematic errors,
while the systematic uncertainties on luminosity and branching ratio are shown as boxes around the data points. The symbols
are plotted at the center of each bin. 
The CMS value  was obtained by integrating the published
$\dd^2 \sigma_{\rm prompt \; J/\psi } / \dd p_{\rm t}\dd y$ data measured 
for $1.6<\left| y \right|<2.4$~\cite{Kha10}.
The results obtained 
by LHCb~\cite{Aai11} and CMS
are reflected with respect to $y = 0$ (open symbols).  
}
\label{fig:4b}
\end{figure}

The differential distribution 
$\frac{\dd^2 \sigma_{\rm prompt \; J/\psi}}{\dd p_{\rm t}\dd y}$ is shown as a function of \pt\ 
in figure~\ref{fig:4} and   
$\frac{\dd \sigma_{\rm prompt \; J/\psi}}{ \dd y} $  is plotted in 
figure~\ref{fig:4b}.  
The numerical values are summarized in table~\ref{tab:2}. 
In figure~\ref{fig:4} the statistical and all systematic errors are added in quadrature for better visibility,  
while in figure~\ref{fig:4b} the error bar shows the quadratic sum of statistical and systematic errors, 
except for    
the 3.5\% systematic uncertainty on luminosity and the 1\% on the branching ratio ($BR$), which are 
added in quadrature and shown as box.  
The results shown in figures~\ref{fig:4} and \ref{fig:4b} 
assume unpolarized \jpsi\ production.
Systematic uncertainties due to the unknown \jpsi\ polarization
are not shown. 
Results by the 
CMS~\cite{Kha10,Chat12},
LHCb~\cite{Aai11} 
and ATLAS~\cite{Aad11} 
Collaborations are shown for comparison. Also for these
data the uncertainties due to 
luminosity 
and to the $BR$  
are shown separately (boxes) in figure~\ref{fig:4b}, while the error bars represent the statistical and the 
other sources of systematic uncertainties added in quadrature.  

The ALICE $\frac{\dd^2 \sigma_{\rm prompt \; J/\psi}}{\dd y \dd p_{\rm t}}$ measurement at 
mid-rapidity (left panel of figure~\ref{fig:4}) is  complementary to the 
data of CMS, available 
for $|y| < 0.9$ and \pt~$>$~8~GeV/$c$, 
and ATLAS, which covers the region $|y| < 0.75$ and \pt~$>$~7~GeV/$c$.
In the right panel of figure~\ref{fig:4}, the ALICE results are compared to next-to-leading  
order (NLO)  non-relativistic QCD (NRQCD) theoretical  
calculations by M.~Butensch\"on and B.A.~Kniehl~\cite{But11} and   Y.-Q.~Ma et al.~\cite{Ma11}.  
Both calculations include color-singlet (CS), color-octet (CO), 
and heavier charmonium feed-down contributions. For one of the two models (M.~Butensch\"on and B.A.~Kniehl) 
the partial results 
with only the CS contribution are also shown.   
The comparison suggests that the CO processes are indispensable to describe the data also  at low \pt. 
The results are also compared to the model of V.A.~Saleev et al.~\cite{Sal12}, which includes the 
contribution of partonic sub-processes involving t-channel parton 
 exchanges and provides a prediction down to $p_{\rm t}=0$.  

The ALICE result for $\frac{\dd \sigma_{\rm prompt \; J/\psi}}{\dd y}$ (figure~\ref{fig:4b}), 
which  equals  
\[
\frac{\dd \sigma_{\rm prompt \; J/\psi}}{\dd y} = 
5.89 \pm 0.60 {\rm (stat.)} ^{+0.88}_{-0.90} {\rm (syst.)} ^{+ 0.03}_{-0.01} {\rm (extr.)} ^{+ 1.01 {\rm (\lambda_{HE}=1)}}_{-0.99 {\rm (\lambda_
{HE}=-1)}} \, {\rm \mu b},  
\]
was obtained 
by subtracting
from the inclusive \jpsi\ cross section measured for $p_{\rm t}>0$  
that of \jpsi\ coming from b-hadron decays.
The latter was determined, as discussed in the next section, by extrapolating the cross section
from the measured region down to $p_{\rm t}>0$ using 
an implementation of pQCD calculations at fixed order with next-to leading-log 
resummation (FONLL)~\cite{Cac04}.  
The extrapolation uncertainty 
is negligible with respect to the 
other systematic uncertainties. 
In figure~\ref{fig:4b} the CMS and LHCb results for the 
rapidity bins where the \pt\ coverage extends down to zero 
were selected.  
For CMS, the value
for $1.6 < |y| < 2.4$ was obtained by integrating the published
$\dd^2 \sigma_{\rm prompt \; J/\psi } / \dd p_{\rm t}\dd y$ data~\cite{Kha10}.  
The ALICE data point at mid-rapidity  
complements 
the other LHC measurements of prompt \jpsi\ production cross section as a function of rapidity. 
It is worth noting that 
the uncertainties of the data sets of the three experiments are uncorrelated, except for that (negligible) of the $BR$, 
while within the same experiment most of the systematic uncertainties are correlated. 
The prediction of the model by V.A.~Saleev et al.~\cite{Sal12} 
at mid-rapidity provides $\frac{{\rm d}\sigma_{\rm prompt \, J/\psi}}{{\rm d}y} = 7.8 ^{+9.7}_{-4.5} \; {\rm \mu b}$, which, 
within the large band of theoretical uncertainties, is in agreement with our measurement.  

\subsection{Beauty hadron production}
\label{beauty}
The cross section of \jpsi\ from b-hadrons decay was obtained as 
$\sigma_{{\rm J/\psi} \leftarrow {\rm h_B} }=f_{\rm B}\cdot \sigma_{\rm J/\psi}$.
In the measured region it is  
\[
\sigma_{{\rm J/\psi} \leftarrow {\rm h_B} }(p_{\rm t}>1.3\, {\rm GeV/}c , \; |y|<0.9) = 
1.46 \pm 0.38 {\rm (stat.)} ^{+0.26}_{-0.32} {\rm (syst.)} \, {\rm \mu b}. 
\]
This measurement can be compared to theoretical calculations based 
on the factorization approach. 
In particular, the prediction of the FONLL~\cite{Cac04}, which 
describes well the beauty production at Tevatron energy, 
provides~\cite{CacPr}  
$
1.33 ^{+0.59}_{-0.48}  \, {\rm \mu b}, 
$ 
in good agreement with the measurement. For this calculation CTEQ6.6 parton 
distribution functions~\cite{Nad08} were used and the theoretical uncertainty was obtained by varying 
the factorization and 
renormalization scales, $\mu_{\rm F}$ and $\mu_{\rm R}$, 
independently in the ranges $0.5 < \mu_{\rm F}/m_{\rm t} < 2$, $0.5 < \mu_{\rm R}/m_{\rm t} < 2$, 
with the constraint $0.5 < \mu_{\rm F}/\mu_{\rm R} < 2$, where 
$m_{\rm t}=\sqrt{p_{\rm t}^2+ m_{\rm b}^2}$. The beauty quark mass was varied within 
$4.5 <m_{\rm b}< 5.0$~GeV/$c^2$.  

\begin{figure}[tb]
\centering{\includegraphics[width=.60\textwidth]{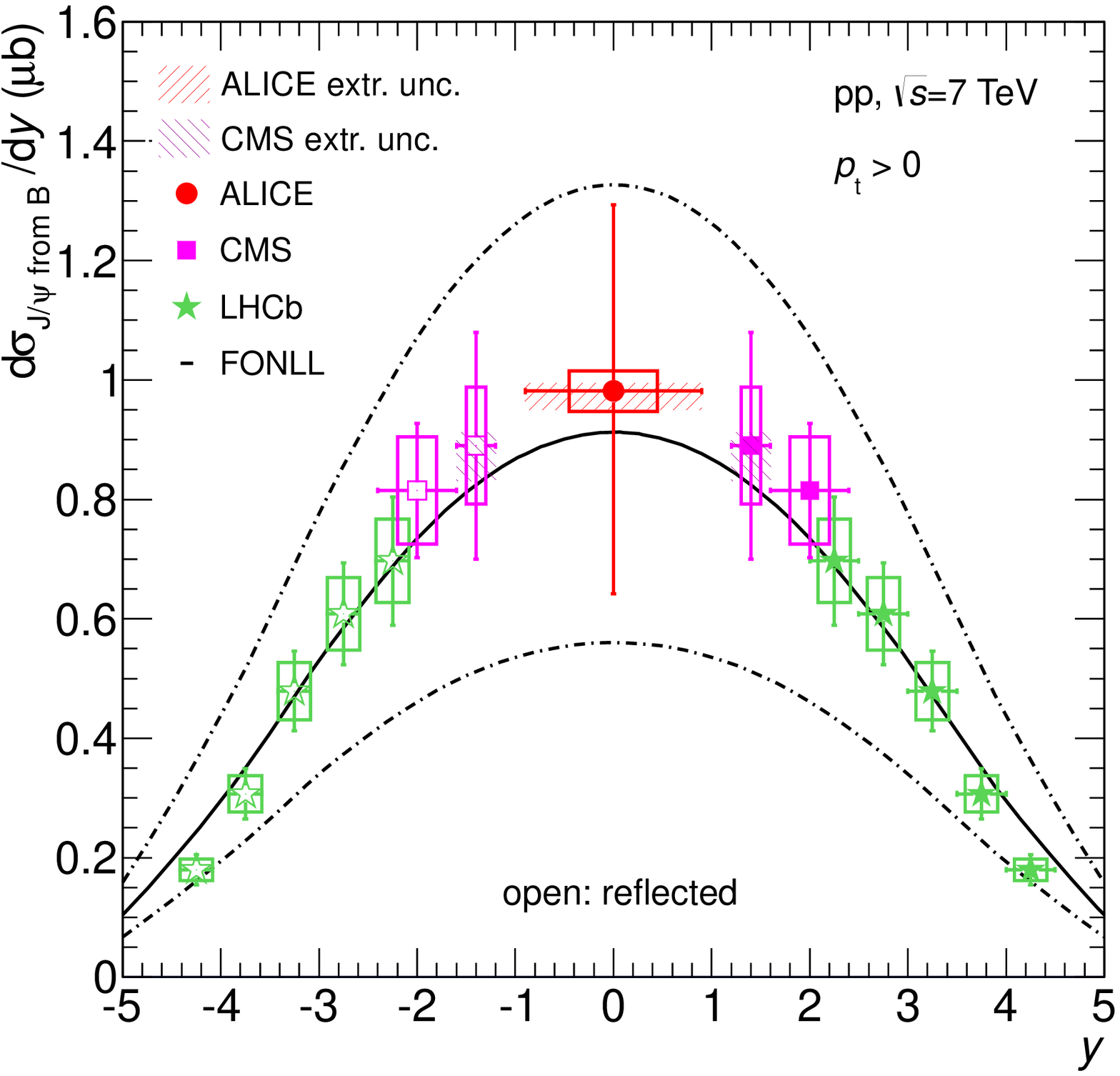}}
\caption{
$\frac{\dd \sigma_{\rm J/\psi \, from \, B}}{\dd y}$\ as a function of $y$.  
The error bars represent the quadratic sum of the statistical and systematic errors,  
while the systematic uncertainties on luminosity and branching ratio are shown as boxes. 
The systematic uncertainties on the extrapolation to \pt~=~0 are indicated by the slashed areas.  
The CMS values  were obtained by integrating the published
$\dd^2 \sigma_{\rm J/\psi\, from \, B } / \dd p_{\rm t}\dd y$ data 
measured for $1.2 <\left| y  \right|<1.6$  and $1.6 <\left| y  \right|<2.4$~\cite{Kha10}.
The results obtained in the forward region by LHCb~\cite{Aai11}   
are reflected with respect to $y = 0$ (open symbols).  
The FONLL calculation~\cite{Cac04,CacPr}  (and its uncertainty) is represented by solid (dashed) lines.
}
\label{fig:5}
\end{figure}

The same FONLL calculations were used to extrapolate the cross section of non-prompt \jpsi\ 
down to \pt\ equal to zero. The 
extrapolation factor, which is equal to $1.212^{+0.016}_{-0.038}$, 
was computed as the ratio of the cross section for $p_{\rm t}^{\rm J/\psi}>0$ 
and $|y_{\rm J/\psi}|<0.9$ to that in the measured region ($p_{\rm t}^{\rm J/\psi}>1.3$~GeV/$c$ and $|y_{\rm J/\psi}|<0.9$). 
Using the PYTHIA event generator with Perugia-0 tuning instead of FONLL provides an extrapolation factor of  
$1.156$. 
The measured cross section corresponds thus to about 80\% of the 
\pt-integrated   
cross section at mid-rapidity.
Dividing 
by the rapidity range $\Delta y=1.8$ one obtains    
\[
\frac{\dd \sigma_{\rm J/\psi \leftarrow {\rm h_B}}}{\dd y} =  
{\rm 0.98 \pm 0.26 \,(stat.) ^{+0.18}_{-0.22}  \, (syst.) ^{+0.01}_{-0.03}\, (extr.)} \; {\rm \mu b}.  
\] 
In figure~\ref{fig:5} this measurement is plotted together with the LHCb~\cite{Aai11} 
and CMS~\cite{Kha10} data at forward rapidity.   
For CMS the values for $1.2<|y|<1.6$ and $1.6<|y|<2.4$  were obtained by integrating 
the published $\dd^2 \sigma_{\rm J/\psi \leftarrow {\rm h_B}}/\dd p_{\rm t} \dd y$ 
data~\cite{Kha10}; the value for $1.2<|y|<1.6$ was also  extrapolated from $p_{\rm t}^{\rm min}=2.0$~GeV/$c$ 
to $p_{\rm t}=0$, with the approach based on the FONLL calculations as previously described. The extrapolation 
uncertainties are shown in figure~\ref{fig:5} as the slashed areas. 
The central FONLL prediction and its uncertainty band are 
also shown. A good agreement between data and theory is observed. 

\begin{figure}[tb]
\centering{\includegraphics[width=.60\textwidth]{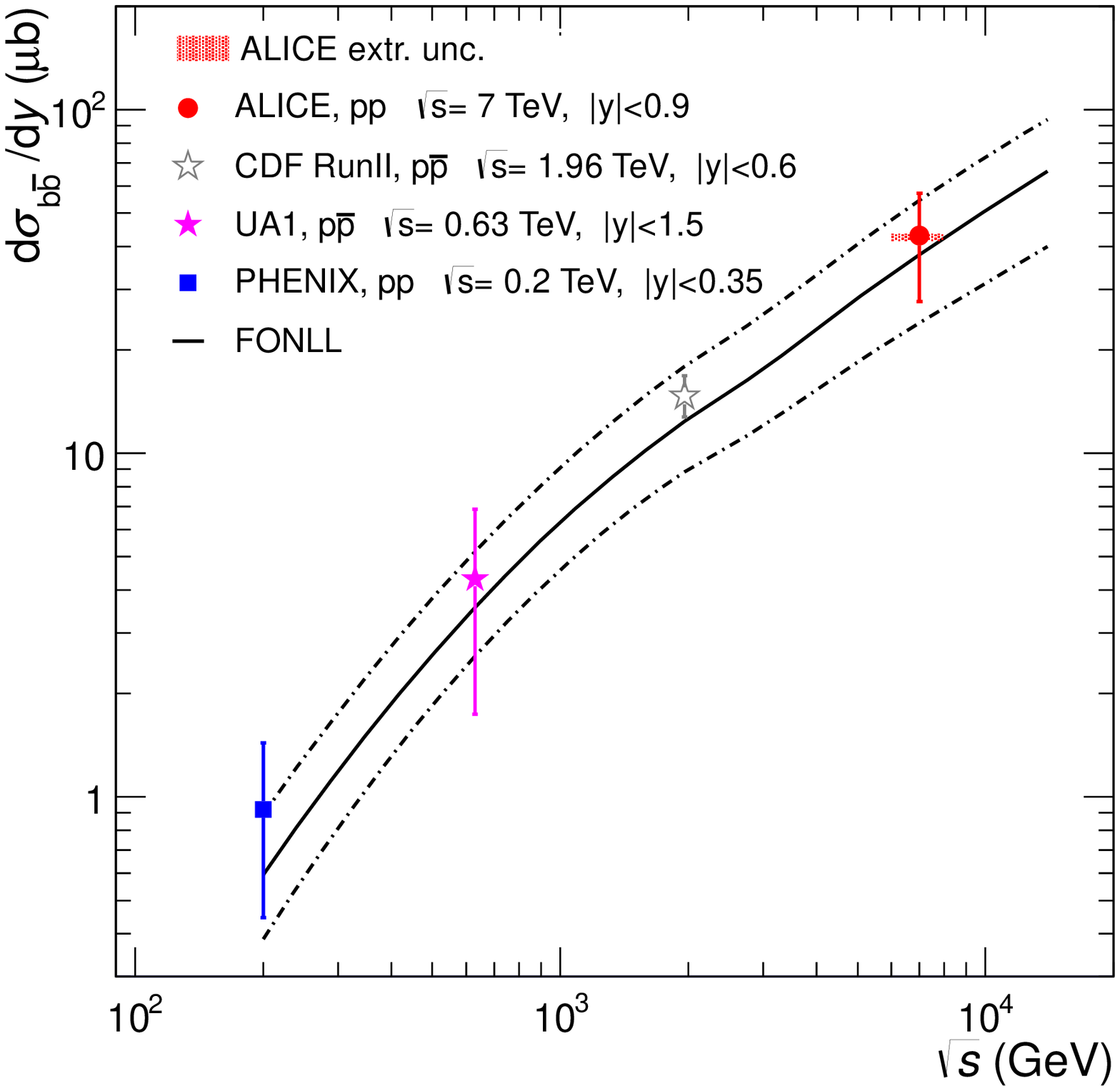}}
\caption{ $\dd \sigma_{\rm b \bar{b}}/\dd y$  at mid rapidity as a function of $\sqrt{s}$ in pp 
(PHENIX~\cite{Ada09} and ALICE results) and ${\rm p\bar{p}}$  
(UA1~\cite{Alb91} and CDF~\cite{Abe95} results) collisions.
The FONLL calculation~\cite{Cac04,CacPr}  (and its uncertainty) is represented by solid (dashed) lines.
}
\label{fig:6}
\end{figure}

A similar procedure  was used to derive the ${\rm b\bar{b}}$ quark-pair production cross section      
%
%
\begin{equation}
\frac{\dd \sigma_{\rm b\bar{b}}}{\dd y} 
 = 
\frac{\dd \sigma_{\rm b\bar{b}}^{\rm theory}}{\dd y} \times
\frac{  
\sigma_{\rm J/\psi \leftarrow {\rm h_B}}(p_{\rm t}^{\rm J/\psi}>1.3 \, 
                          {\rm GeV}/c{\rm ,} \; |y_{\rm J/\psi}|<0.9 )
}{ 
\sigma_{\rm J/\psi \leftarrow {\rm h_B}}^{\rm theory}(p_{\rm t}^{\rm J/\psi}>1.3 \, 
                          {\rm GeV}/c{\rm ,} \; |y_{\rm J/\psi}|<0.9 )   
} ,
\label{eq:extr}
\end{equation}
where the average branching fraction of inclusive b-hadron decays to \jpsi\ measured at LEP~\cite{Abr94,Adr93,Bus92}, 
$BR(\rm{h_b\rightarrow J/\psi +X})= (1.16 \pm 0.10)\%$, was used in the computation of $\sigma_{\rm J/\psi \leftarrow {\rm h_B}}^{\rm theory}$. 
The extrapolation with the  FONLL calculations provides  
\[ \frac{\dd \sigma_{\rm b \bar{b}}}{\dd y} = 43 \pm 11 \, ({\rm stat.})  ^{+9}_{-10} ({\rm syst.}) ^{+0.6}_{-1.5} ({\rm extr.}) \; {\rm \mu b}. \] 
%
Using the PYTHIA event generator with Perugia-0 tuning (with the EvtGen package to describe the particle decays) instead of FONLL results in  
a central value of 
40.4 ($40.9$)~${\rm \mu b}$.  
A compilation of measurements of $\dd \sigma_{\rm b \bar{b}}/\dd y$ at mid-rapidity is plotted in 
figure~\ref{fig:6} as a function of $\sqrt{s}$, with superimposed FONLL predictions. 

Finally, the total ${\rm b\bar{b}}$ cross section was obtained as   
\begin{equation}
\sigma({\rm pp \rightarrow b\bar{b}}+X) = \alpha_{4\pi} 
   \frac{\sigma_{\rm J/\psi \leftarrow h_B}(p_{\rm t}^{\rm J/\psi}>1.3 \, 
                          {\rm GeV}/c{\rm ,} \; |y_{\rm J/\psi}|<0.9 )}
   {2 \cdot \, BR({\rm h_b\rightarrow J/\psi +X})},
\end{equation}
where $\alpha_{4\pi}$ is the ratio between the yield of \jpsi\ mesons  
(from the decay of b-hadrons) in the full phase space and the yield 
in the measured  region $|y_{\rm J/\psi}| < 0.9$ and $p_{\rm t}^{\rm J/\psi}>1.3$~GeV/$c$.  
The FONLL calculations provide $\alpha_{4\pi}=4.49^{+0.12}_{-0.10}$, which produces 
$\sigma({\rm pp \rightarrow b\bar{b}}+X) = 
 282 \pm 74 ({\rm stat.}) ^{+58}_{-68} ({\rm syst.}) ^{+8}_{-7} ({\rm extr.}) \; {\rm \mu b}$.   
The extrapolation factor $\alpha_{4\pi}$
was also estimated using PYTHIA with Perugia-0 tuning and found to be 
$\alpha_{4\pi}^{\rm PYTHIA}=4.20$.   
This measurement is in good agreement with those of the LHCb experiment, namely $288 \pm 4 ({\rm stat.}) \pm 48 ({\rm syst.})\; {\rm \mu b}$ and 
$284 \pm 20 ({\rm stat.}) \pm 49 ({\rm syst.}) \; {\rm \mu b}$, which 
were based on the  
measured cross sections determined in the forward rapidity range 
from  b-hadron decays into ${\rm J/\psi} X$ and ${\rm D^0 \mu \nu} X$, respectively~\cite{Aai11,Aai10}. 
\section {Summary}
Results on the production cross section of  prompt \jpsi\ and \jpsi\ from the decay of b-hadrons at 
mid-rapidity in pp collisions at $\sqrt{s}=7$~TeV have been presented. 
The measured cross sections have been compared to theoretical predictions based on QCD  
and results from other experiments. Prompt \jpsi\ production is well described 
by NLO NRQCD models that include color-octet processes.   
The cross section of \jpsi\ from b-hadron decays 
is in good agreement with the 
FONLL prediction, based on perturbative QCD.   
The ALICE results at mid-rapidity, covering a lower \pt\ region down to $p_{\rm t}=1.3$~GeV/$c$, 
are complementary to those of the ATLAS and CMS 
experiments, which are available for \jpsi\ \pt\ above 
6.5~GeV/$c$. 
Using the shape of the \pt\ and $y$ distributions of b-quarks predicted by 
FONLL calculations, the mid-rapidity $\dd \sigma/\dd y$ and the total  
production cross section of  ${\rm b\bar{b}}$ pairs  
were determined. 

%
\newenvironment{acknowledgement}{\relax}{\relax}
\begin{acknowledgement}
\section{Acknowledgements}
The ALICE collaboration would like to thank all its engineers and technicians for their invaluable contributions to the construction of the experiment and the CERN accelerator teams for the outstanding performance of the LHC complex.
\\
The ALICE collaboration would like to thank M.~Butensch\"on and B.A.~Kniehl,
Y.-Q. Ma, K. Wang and K.T. Chao,
and V.A.~Saleev, M.A.~Nefedov and A.V.~Shipilova  
 for providing their 
theoretical computations of
the
production cross section of prompt \jpsi, and M. Cacciari
for predictions in the FONLL scheme. \\
The ALICE collaboration acknowledges the following funding agencies for their support in building and
running the ALICE detector:
 \\
Calouste Gulbenkian Foundation from Lisbon and Swiss Fonds Kidagan, Armenia;
 \\
Conselho Nacional de Desenvolvimento Cient\'{\i}fico e Tecnol\'{o}gico (CNPq), Financiadora de Estudos e Projetos (FINEP),
Funda\c{c}\~{a}o de Amparo \`{a} Pesquisa do Estado de S\~{a}o Paulo (FAPESP);
 \\
National Natural Science Foundation of China (NSFC), the Chinese Ministry of Education (CMOE)
and the Ministry of Science and Technology of China (MSTC);
 \\
Ministry of Education and Youth of the Czech Republic;
 \\
Danish Natural Science Research Council, the Carlsberg Foundation and the Danish National Research Foundation;
 \\
The European Research Council under the European Community's Seventh Framework Programme;
 \\
Helsinki Institute of Physics and the Academy of Finland;
 \\
French CNRS-IN2P3, the `Region Pays de Loire', `Region Alsace', `Region Auvergne' and CEA, France;
 \\
German BMBF and the Helmholtz Association;
\\
General Secretariat for Research and Technology, Ministry of
Development, Greece;
\\
Hungarian OTKA and National Office for Research and Technology (NKTH);
 \\
Department of Atomic Energy and Department of Science and Technology of the Government of India;
 \\
Istituto Nazionale di Fisica Nucleare (INFN) of Italy;
 \\
MEXT Grant-in-Aid for Specially Promoted Research, Ja\-pan;
 \\
Joint Institute for Nuclear Research, Dubna;
 \\
National Research Foundation of Korea (NRF);
 \\
CONACYT, DGAPA, M\'{e}xico, ALFA-EC and the HELEN Program (High-Energy physics Latin-American--European Network);
 \\
Stichting voor Fundamenteel Onderzoek der Materie (FOM) and the Nederlandse Organisatie voor Wetenschappelijk Onderzoek (NWO), Netherlands;
 \\
Research Council of Norway (NFR);
 \\
Polish Ministry of Science and Higher Education;
 \\
National Authority for Scientific Research - NASR (Autoritatea Na\c{t}ional\u{a} pentru Cercetare \c{S}tiin\c{t}ific\u{a} - ANCS);
 \\
Federal Agency of Science of the Ministry of Education and Science of Russian Federation, International Science and
Technology Center, Russian Academy of Sciences, Russian Federal Agency of Atomic Energy, Russian Federal Agency for Science and Innovations and CERN-INTAS;
 \\
Ministry of Education of Slovakia;
 \\
Department of Science and Technology, South Africa;
 \\
CIEMAT, EELA, Ministerio de Educaci\'{o}n y Ciencia of Spain, Xunta de Galicia (Conseller\'{\i}a de Educaci\'{o}n),
CEA\-DEN, Cubaenerg\'{\i}a, Cuba, and IAEA (International Atomic Energy Agency);
 \\
Swedish Research Council (VR) and Knut $\&$ Alice Wallenberg
Foundation (KAW);
 \\
Ukraine Ministry of Education and Science;
 \\
United Kingdom Science and Technology Facilities Council (STFC);
 \\
The United States Department of Energy, the United States National
Science Foundation, the State of Texas, and the State of Ohio.
\end{acknowledgement}
\newpage
%
%
\appendix
\section{The ALICE Collaboration}
\label{app:collab}

\begingroup
\small
\begin{flushleft}
B.~Abelev\Irefn{org1234}\And
J.~Adam\Irefn{org1274}\And
D.~Adamov\'{a}\Irefn{org1283}\And
A.M.~Adare\Irefn{org1260}\And
M.M.~Aggarwal\Irefn{org1157}\And
G.~Aglieri~Rinella\Irefn{org1192}\And
A.G.~Agocs\Irefn{org1143}\And
A.~Agostinelli\Irefn{org1132}\And
S.~Aguilar~Salazar\Irefn{org1247}\And
Z.~Ahammed\Irefn{org1225}
% ...
\renewcommand\labelenumi{\textsuperscript{\theenumi}~}
\section*{Affiliation notes}
\renewcommand\theenumi{\roman{enumi}}
\begin{Authlist}
\item \Adef{M.V.Lomonosov Moscow State University, D.V.Skobeltsyn Institute of Nuclear Physics, Moscow, Russia}Also at: M.V.Lomonosov Moscow State University, D.V.Skobeltsyn Institute of Nuclear Physics, Moscow, Russia
\item \Adef{Institute of Nuclear Sciences, Belgrade, Serbia}Also at: "Vin\v{c}a" Institute of Nuclear Sciences, Belgrade, Serbia
\end{Authlist}
\section*{Collaboration Institutes}
\renewcommand\theenumi{\arabic{enumi}~}
\begin{Authlist}
\item \Idef{org1279}Benem\'{e}rita Universidad Aut\'{o}noma de Puebla, Puebla, Mexico
\item \Idef{org1220}Bogolyubov Institute for Theoretical Physics, Kiev, Ukraine
\item \Idef{org1262}Budker Institute for Nuclear Physics, Novosibirsk, Russia
\item \Idef{org1292}California Polytechnic State University, San Luis Obispo, California, United States
\item \Idef{org14939}Centre de Calcul de l'IN2P3, Villeurbanne, France
\item \Idef{org1197}Centro de Aplicaciones Tecnol\'{o}gicas y Desarrollo Nuclear (CEADEN), Havana, Cuba
\item \Idef{org1242}Centro de Investigaciones Energ\'{e}ticas Medioambientales y Tecnol\'{o}gicas (CIEMAT), Madrid, Spain
\item \Idef{org1244}Centro de Investigaci\'{o}n y de Estudios Avanzados (CINVESTAV), Mexico City and M\'{e}rida, Mexico
\item \Idef{org1335}Centro Fermi -- Centro Studi e Ricerche e Museo Storico della Fisica ``Enrico Fermi'', Rome, Italy
\item \Idef{org17347}Chicago State University, Chicago, United States
\item \Idef{org1288}Commissariat \`{a} l'Energie Atomique, IRFU, Saclay, France
\item \Idef{org1294}Departamento de F\'{\i}sica de Part\'{\i}culas and IGFAE, Universidad de Santiago de Compostela, Santiago de Compostela, Spain
\item \Idef{org1106}Department of Physics Aligarh Muslim University, Aligarh, India
\item \Idef{org1121}Department of Physics and Technology, University of Bergen, Bergen, Norway
\item \Idef{org1162}Department of Physics, Ohio State University, Columbus, Ohio, United States
\item \Idef{org1300}Department of Physics, Sejong University, Seoul, South Korea
\item \Idef{org1268}Department of Physics, University of Oslo, Oslo, Norway
\item \Idef{org1145}Dipartimento di Fisica dell'Universit\`{a} and Sezione INFN, Cagliari, Italy
\item \Idef{org1270}Dipartimento di Fisica dell'Universit\`{a} and Sezione INFN, Padova, Italy
\item \Idef{org1315}Dipartimento di Fisica dell'Universit\`{a} and Sezione INFN, Trieste, Italy
\item \Idef{org1132}Dipartimento di Fisica dell'Universit\`{a} and Sezione INFN, Bologna, Italy
\item \Idef{org1285}Dipartimento di Fisica dell'Universit\`{a} `La Sapienza' and Sezione INFN, Rome, Italy
\item \Idef{org1154}Dipartimento di Fisica e Astronomia dell'Universit\`{a} and Sezione INFN, Catania, Italy
\item \Idef{org1290}Dipartimento di Fisica `E.R.~Caianiello' dell'Universit\`{a} and Gruppo Collegato INFN, Salerno, Italy
\item \Idef{org1312}Dipartimento di Fisica Sperimentale dell'Universit\`{a} and Sezione INFN, Turin, Italy
\item \Idef{org1103}Dipartimento di Scienze e Innovazione Tecnologica dell'Universit\`{a} del Piemonte Orientale and Gruppo Collegato INFN, Alessandria, Italy
\item \Idef{org1114}Dipartimento Interateneo di Fisica `M.~Merlin' and Sezione INFN, Bari, Italy
\item \Idef{org1237}Division of Experimental High Energy Physics, University of Lund, Lund, Sweden
\item \Idef{org1192}European Organization for Nuclear Research (CERN), Geneva, Switzerland
\item \Idef{org1227}Fachhochschule K\"{o}ln, K\"{o}ln, Germany
\item \Idef{org1122}Faculty of Engineering, Bergen University College, Bergen, Norway
\item \Idef{org1136}Faculty of Mathematics, Physics and Informatics, Comenius University, Bratislava, Slovakia
\item \Idef{org1274}Faculty of Nuclear Sciences and Physical Engineering, Czech Technical University in Prague, Prague, Czech Republic
\item \Idef{org1229}Faculty of Science, P.J.~\v{S}af\'{a}rik University, Ko\v{s}ice, Slovakia
\item \Idef{org1184}Frankfurt Institute for Advanced Studies, Johann Wolfgang Goethe-Universit\"{a}t Frankfurt, Frankfurt, Germany
\item \Idef{org1215}Gangneung-Wonju National University, Gangneung, South Korea
\item \Idef{org1212}Helsinki Institute of Physics (HIP) and University of Jyv\"{a}skyl\"{a}, Jyv\"{a}skyl\"{a}, Finland
\item \Idef{org1203}Hiroshima University, Hiroshima, Japan
\item \Idef{org1329}Hua-Zhong Normal University, Wuhan, China
\item \Idef{org1254}Indian Institute of Technology, Mumbai, India
\item \Idef{org36378}Indian Institute of Technology Indore (IIT), Indore, India
\item \Idef{org1266}Institut de Physique Nucl\'{e}aire d'Orsay (IPNO), Universit\'{e} Paris-Sud, CNRS-IN2P3, Orsay, France
\item \Idef{org1277}Institute for High Energy Physics, Protvino, Russia
\item \Idef{org1249}Institute for Nuclear Research, Academy of Sciences, Moscow, Russia
\item \Idef{org1320}Nikhef, National Institute for Subatomic Physics and Institute for Subatomic Physics of Utrecht University, Utrecht, Netherlands
\item \Idef{org1250}Institute for Theoretical and Experimental Physics, Moscow, Russia
\item \Idef{org1230}Institute of Experimental Physics, Slovak Academy of Sciences, Ko\v{s}ice, Slovakia
\item \Idef{org1127}Institute of Physics, Bhubaneswar, India
\item \Idef{org1275}Institute of Physics, Academy of Sciences of the Czech Republic, Prague, Czech Republic
\item \Idef{org1139}Institute of Space Sciences (ISS), Bucharest, Romania
\item \Idef{org27399}Institut f\"{u}r Informatik, Johann Wolfgang Goethe-Universit\"{a}t Frankfurt, Frankfurt, Germany
\item \Idef{org1185}Institut f\"{u}r Kernphysik, Johann Wolfgang Goethe-Universit\"{a}t Frankfurt, Frankfurt, Germany
\item \Idef{org1177}Institut f\"{u}r Kernphysik, Technische Universit\"{a}t Darmstadt, Darmstadt, Germany
\item \Idef{org1256}Institut f\"{u}r Kernphysik, Westf\"{a}lische Wilhelms-Universit\"{a}t M\"{u}nster, M\"{u}nster, Germany
\item \Idef{org1246}Instituto de Ciencias Nucleares, Universidad Nacional Aut\'{o}noma de M\'{e}xico, Mexico City, Mexico
\item \Idef{org1247}Instituto de F\'{\i}sica, Universidad Nacional Aut\'{o}noma de M\'{e}xico, Mexico City, Mexico
\item \Idef{org23333}Institut of Theoretical Physics, University of Wroclaw
\item \Idef{org1308}Institut Pluridisciplinaire Hubert Curien (IPHC), Universit\'{e} de Strasbourg, CNRS-IN2P3, Strasbourg, France
\item \Idef{org1182}Joint Institute for Nuclear Research (JINR), Dubna, Russia
\item \Idef{org1143}KFKI Research Institute for Particle and Nuclear Physics, Hungarian Academy of Sciences, Budapest, Hungary
\item \Idef{org1199}Kirchhoff-Institut f\"{u}r Physik, Ruprecht-Karls-Universit\"{a}t Heidelberg, Heidelberg, Germany
\item \Idef{org20954}Korea Institute of Science and Technology Information, Daejeon, South Korea
\item \Idef{org1160}Laboratoire de Physique Corpusculaire (LPC), Clermont Universit\'{e}, Universit\'{e} Blaise Pascal, CNRS--IN2P3, Clermont-Ferrand, France
\item \Idef{org1194}Laboratoire de Physique Subatomique et de Cosmologie (LPSC), Universit\'{e} Joseph Fourier, CNRS-IN2P3, Institut Polytechnique de Grenoble, Grenoble, France
\item \Idef{org1187}Laboratori Nazionali di Frascati, INFN, Frascati, Italy
\item \Idef{org1232}Laboratori Nazionali di Legnaro, INFN, Legnaro, Italy
\item \Idef{org1125}Lawrence Berkeley National Laboratory, Berkeley, California, United States
\item \Idef{org1234}Lawrence Livermore National Laboratory, Livermore, California, United States
\item \Idef{org1251}Moscow Engineering Physics Institute, Moscow, Russia
\item \Idef{org1140}National Institute for Physics and Nuclear Engineering, Bucharest, Romania
\item \Idef{org1165}Niels Bohr Institute, University of Copenhagen, Copenhagen, Denmark
\item \Idef{org1109}Nikhef, National Institute for Subatomic Physics, Amsterdam, Netherlands
\item \Idef{org1283}Nuclear Physics Institute, Academy of Sciences of the Czech Republic, \v{R}e\v{z} u Prahy, Czech Republic
\item \Idef{org1264}Oak Ridge National Laboratory, Oak Ridge, Tennessee, United States
\item \Idef{org1189}Petersburg Nuclear Physics Institute, Gatchina, Russia
\item \Idef{org1170}Physics Department, Creighton University, Omaha, Nebraska, United States
\item \Idef{org1157}Physics Department, Panjab University, Chandigarh, India
\item \Idef{org1112}Physics Department, University of Athens, Athens, Greece
\item \Idef{org1152}Physics Department, University of Cape Town, iThemba LABS, Cape Town, South Africa
\item \Idef{org1209}Physics Department, University of Jammu, Jammu, India
\item \Idef{org1207}Physics Department, University of Rajasthan, Jaipur, India
\item \Idef{org1200}Physikalisches Institut, Ruprecht-Karls-Universit\"{a}t Heidelberg, Heidelberg, Germany
\item \Idef{org1325}Purdue University, West Lafayette, Indiana, United States
\item \Idef{org1281}Pusan National University, Pusan, South Korea
\item \Idef{org1176}Research Division and ExtreMe Matter Institute EMMI, GSI Helmholtzzentrum f\"ur Schwerionenforschung, Darmstadt, Germany
\item \Idef{org1334}Rudjer Bo\v{s}kovi\'{c} Institute, Zagreb, Croatia
\item \Idef{org1298}Russian Federal Nuclear Center (VNIIEF), Sarov, Russia
\item \Idef{org1252}Russian Research Centre Kurchatov Institute, Moscow, Russia
\item \Idef{org1224}Saha Institute of Nuclear Physics, Kolkata, India
\item \Idef{org1130}School of Physics and Astronomy, University of Birmingham, Birmingham, United Kingdom
\item \Idef{org1338}Secci\'{o}n F\'{\i}sica, Departamento de Ciencias, Pontificia Universidad Cat\'{o}lica del Per\'{u}, Lima, Peru
\item \Idef{org1316}Sezione INFN, Trieste, Italy
\item \Idef{org1271}Sezione INFN, Padova, Italy
\item \Idef{org1313}Sezione INFN, Turin, Italy
\item \Idef{org1286}Sezione INFN, Rome, Italy
\item \Idef{org1146}Sezione INFN, Cagliari, Italy
\item \Idef{org1133}Sezione INFN, Bologna, Italy
\item \Idef{org1115}Sezione INFN, Bari, Italy
\item \Idef{org1155}Sezione INFN, Catania, Italy
\item \Idef{org1322}Soltan Institute for Nuclear Studies, Warsaw, Poland
\item \Idef{org36377}Nuclear Physics Group, STFC Daresbury Laboratory, Daresbury, United Kingdom
\item \Idef{org1258}SUBATECH, Ecole des Mines de Nantes, Universit\'{e} de Nantes, CNRS-IN2P3, Nantes, France
\item \Idef{org1304}Technical University of Split FESB, Split, Croatia
\item \Idef{org1168}The Henryk Niewodniczanski Institute of Nuclear Physics, Polish Academy of Sciences, Cracow, Poland
\item \Idef{org17361}The University of Texas at Austin, Physics Department, Austin, TX, United States
\item \Idef{org1173}Universidad Aut\'{o}noma de Sinaloa, Culiac\'{a}n, Mexico
\item \Idef{org1296}Universidade de S\~{a}o Paulo (USP), S\~{a}o Paulo, Brazil
\item \Idef{org1149}Universidade Estadual de Campinas (UNICAMP), Campinas, Brazil
\item \Idef{org1239}Universit\'{e} de Lyon, Universit\'{e} Lyon 1, CNRS/IN2P3, IPN-Lyon, Villeurbanne, France
\item \Idef{org1205}University of Houston, Houston, Texas, United States
\item \Idef{org20371}University of Technology and Austrian Academy of Sciences, Vienna, Austria
\item \Idef{org1222}University of Tennessee, Knoxville, Tennessee, United States
\item \Idef{org1310}University of Tokyo, Tokyo, Japan
\item \Idef{org1318}University of Tsukuba, Tsukuba, Japan
\item \Idef{org21360}Eberhard Karls Universit\"{a}t T\"{u}bingen, T\"{u}bingen, Germany
\item \Idef{org1225}Variable Energy Cyclotron Centre, Kolkata, India
\item \Idef{org1306}V.~Fock Institute for Physics, St. Petersburg State University, St. Petersburg, Russia
\item \Idef{org1323}Warsaw University of Technology, Warsaw, Poland
\item \Idef{org1179}Wayne State University, Detroit, Michigan, United States
\item \Idef{org1260}Yale University, New Haven, Connecticut, United States
\item \Idef{org1332}Yerevan Physics Institute, Yerevan, Armenia
\item \Idef{org15649}Yildiz Technical University, Istanbul, Turkey
\item \Idef{org1301}Yonsei University, Seoul, South Korea
\item \Idef{org1327}Zentrum f\"{u}r Technologietransfer und Telekommunikation (ZTT), Fachhochschule Worms, Worms, Germany
\end{Authlist}
\end{flushleft}
\endgroup
%
%
\end{document}